\documentclass[11pt]{article}
\usepackage{epsfig}

\usepackage{epstopdf}
\usepackage{amsmath}\usepackage{amssymb}
\usepackage{mathrsfs}
\usepackage{graphicx}

\begin{document}
\def \a'{\alpha'}
\baselineskip 0.65 cm
\begin{flushright}
\ \today
\end{flushright}

\begin{center}{\large
{\bf Freeze-in production of Fermionic Dark Matter with Pseudo-scalar and Phenomenological Aspects}} {\vskip 0.5 cm} {\bf ${\rm Seyed~ Yaser~ Ayazi}$$^1$, ${\rm S.~Mahdi~ Firouzabadi}$$^1$ and ${\rm S.~Peyman~ Zakeri}$$^{1,2}$}{\vskip 0.5 cm
}
{\small $^1$$School~ of~ Particles~ and~ Accelerators,~ Institute~ for~ Research~ in~
Fundamental$ $ Sciences~(IPM), P.O.~Box~ 19395-5531, Tehran, Iran$
$^2$$Department~ of~ Physics,~ Yazd~ University,$ $~ P.O. Box~89195-741,~Yazd,~ Iran$}
\end{center}

\begin{abstract}
In this paper, we study freeze-in production of fermionic dark matter with a pseudo scalar as the mediator between dark sector and Standard Model (SM). While the fermionic DM is non-thermal, we will explain two scenarios in which production of pseudo-scalar particles are either thermal or non thermal. We'll present elaborate discussion to derive yield quantity and relic density and illustrate these values for the different range of model parameters. We'll investigate constraints on parameters space coming from invisible Higgs decay at LHC. For the case of extremely small couplings and zero mixing between SM Higgs field with pseudo scalar field which pseudo scalar boson can play role of DM, we will justify recent observation of merging galaxies with a case of self-interacting DM. We'll show that influence of DM annihilation in this case, would provide a better fit to the AMS-02 data of positron flux.
\end{abstract}

\section{Introduction}
While the existence of dark matter (DM) has been confirmed by several lines of evidence, such as galactic rotation curves, gravitational lensing, observations of merging galaxies and etc, its detailed properties remain unknown. A most attractive assumption is that DM consists of Weakly Interactive Massive Particles (WIMPs) with mass of a few hundred $\rm GeV$. In this scenario, it is supposed that the  relic density of these particles have been produced via freeze-out mechanism in the early Universe \cite{Freez-out}. The possibility of detecting DM WIMPs has been searched by direct detection and indirect experiments. So far, we have not seen a clear signal to approve WIMP scenario in these experiments. While WIMPs cannot be excluded, we can think about other possible scenarios for the production of DM particles. Due to null results of direct detection experiments, models of non-interacting DM or feebly interacting DM hase been favored lately\cite{non-interacting DM}. One of the main  alternative idea for WIMPs DM with freeze-out production is non-thermal production of DM particles \cite{Hall:2009bx}. In this scenario, it is supposed that there is a set of bath particles in thermal equilibrium and a Feebly Interacting Massive Particle (FIMP) which never enters the thermal equilibrium. FIMPs are mostly produced via a mechanism that is called freeze-in. In this mechanism, the initial abundance of FIMPs assumed to be negligible. Since FIMP coupling is small, they are produced slowly and gradually by decay or collision of other bath particles. For this reason, phenomenological impacts of FIMPs in direct detection of DM and colliders are hardly detectable \cite{Hall:2009bx}.

In this paper, we extend SM by addition of a Dirac fermion which is singlet under SM gauge group symmetry. This field is coupled to SM particles via a pseudo scalar. In the following, we focus on providing an overview of freeze-in mechanism of fermionic DM with pseudo scalar as a mediator. We should mention that different aspects of fermionic dark matter with scalar mediator have been discussed in the literature before \cite{fermionic DM}. In particular \cite{Klasen:2013ypa} has studied freeze-in production of fermionic DM with a scalar mediator. We will study evolution of relic denisty for both fermionic DM and pseudo scalar as the universe cools down.
While fermionic DM acts like a FIMP, the pseudo scalar can behave either as a FIMP or WIMP in different regions of parameters space.

 This paper is organized as follows: in section 2, we introduce our model in which a Dirac fermion plays the role of DM. In
section~3, we introduce the main equations to be solved and present our results for DM abundance and relic density of DM in two scenarios in which pseudo scalar mediator produce thermally and non thermally. In section~4, we will study  phenomenological aspects of singlet fermionic DM, pseudo scalar mediator and constraints on its parameters space.
The conclusions are given in section~5. The decay rate and cross section formulae for fermionic DM and pseudo scalar annihilation are summarized in the Appendix.

\section{The Model}
In this model, apart from the SM Higgs doublet, we introduce a Dirac fermion $\chi$ and a real pseudo scalar $\varphi$. The new fields are singlet under SM gauge groups and the DM candidate (fermion field) is charged under a global $U(1)$ symmetry and all SM fields are singlet under the global symmetry. Also kinetic mixing between SM $U(1)$ gauge boson and Dark sector gauge boson has been neglected. There are vast range of studies have been presented on models with such mixing and phenomenological constraints from astronomical and collider experiments on mixing parameter \cite{KM}. This assumption guarantees stability of DM candidate $\chi$ since there isn't any mixing between SM fermions and $\chi$. The only renormalizable interactions of DM candidate $\chi$ is $\varphi\overline{\chi}\gamma^5\chi$. The interaction between $\varphi$ and the SM Higgs boson can provide a link between SM particles and DM sector. The Lagrangian of the model consists the following parts:
\begin{equation}
{\cal L} = {\cal L_{\text{SM}}}+{\cal L}_{\chi,\varphi}+{\cal L}_{\text{int}}-V(\varphi,H)\,.
\end{equation}
The part of the Lagrangian involving the dark matter $\chi$ and pseudo scalar $\varphi$ is given by:
\begin{equation}
{\cal L}_{\chi,\varphi} =  i\bar{\chi}({\not}\partial-m_{DM})\chi+\frac{1}{2}(\partial_{\mu}\varphi)^2\,.
\end{equation}
The scalar potential is modified as follows:
\begin{equation}
V(\varphi,H)=-\mu^{2}_{H} H^{\dagger}H - \lambda_{H} (H^{\dagger}H)^2- \frac{m^{2}_{0}}{2}\varphi^2 -\frac{\lambda}{24}\varphi^4  \,.
\end{equation}
where $H$ is the SM Higgs doublet which causes the electroweak spontaneous symmetry breaking. The vacuum stability is given by:
\begin{equation}
H = \frac{1}{\sqrt{2}} \left( \begin{array}{c}
                                0  \\
                                v_{H}+\tilde{H}
                       \end{array} \right)\,,
\end{equation}
where $v_{H}$ = 246 GeV. The general form of renormalizable Lagrangian for interaction of pseudo scalar fields $\varphi$, SM particles and DM candidate $\chi$ is given by:
\begin{equation}
{\cal L}_{\text{int}}=-i g_p\varphi \bar{\chi}\gamma^{5}\chi - \lambda_{1} \varphi^2 H^{\dagger}H .
\label{int}
\end{equation}
As we demand that ${\cal L}$ to be CP-invariant, the Lagrangian does not include $\varphi$, $\varphi^3$ and $\varphi H^2$ terms. Knowing that $\varphi$ is pseudo-scalar fields, these terms break parity symmetry and we ignore them. This means, our model has minimal interaction in comparison to models with scalar field mediator \cite{Klasen:2013ypa} which contains such interaction.

In principle, $\varphi$ can acquire a VEV,
\begin{equation}
\varphi =  v_{\varphi} + S\,.
\end{equation}
The $\lambda_1$ term in Eq.~\ref{int} induces a mixing between $S$ and $\tilde{H}$ which gives rise to two scalar mass eigenstates $h$ and $\rho$:
\begin{equation}
h = \sin \theta~S + \cos \theta~\tilde{H}\,, ~~~ \rho = \cos \theta~ S - \sin \theta~ \tilde{H}\,,
\end{equation}
where $\theta$ is the mixing angle and is defined by \cite{Ghorbani:2014qpa}:
\begin{equation}
\theta=Arc\tan[\frac{y}{1+\sqrt{1+y^2}}],~~~~~~~~y=\frac{2\lambda_1v_{\varphi}v_{H}}{\lambda_Hv^2_{H}-\frac{1}{6}\lambda v^2_{\varphi}}.
\end{equation}
The mass eigenvalues of scalar fields can be expressed  by:
 \begin{equation}
m^2_{h,\rho}=(\lambda_Hv^2_{H}+\frac{1}{6}\lambda v^2_{\varphi})\pm(\lambda_Hv^2_{H}-\frac{1}{6}\lambda v^2_{\varphi})(1+\sqrt{1+y^2}).
\end{equation}

Note that the two neutral Higgs-like scalars $h$ and $\rho$ have been given as admixtures of
SM Higgs $\tilde{H}$ and pseudo scalar S. (In \cite{Klasen:2013ypa} mixing between SM Higgs and scalar field arises from $\mu \varphi H^{\dagger}H$ which does not exist in our model).
  We apply upper limits on mixing angle between the doublet and scalar that have been presented in \cite{Higgs mixing}.

 The couplings can be obtained in terms of mixing angle and scalar masses:
\begin{eqnarray}
\lambda_H&=&\frac{m^2_{\rho}\sin^2\theta+m^2_{h}\cos^2\theta}{2v^2_{H}}, ~~~~~~~\lambda=\frac{m^2_{\rho}\cos^2\theta+m^2_{h}\sin^2\theta}{v^2_{\varphi}/3}\nonumber \\
\lambda_1&=&\frac{m^2_{\rho}-m^2_{h}}{4v_{H}v_{\varphi}}\sin^2{2\theta}.
\end{eqnarray}
\section{Numerical results for freeze-in DM Production}
In this section, we study the freeze-in mechanism for production of fermionic DM. We assume that the coupling of fermionic DM to the thermal bath is very small and this means DM is a FIMP. In present scenario, in early Universe all SM fields are in equilibrium and fermionic DM does not reach equilibrium and is never abundant enough to annihilate. For this reason, we suppose initial abundance of fermionic DM is negligible so that we may set $f_{\chi}=0$. However pseudo scalar $\rho$ can reach equilibrium when there is a large enough coupling with Higgs field.
In the following, we study fermionic DM production in the phenomenologically favorable cases. In the first case, two newly added fields $\chi$ and $\rho$ are FIMP. In the second case, the fermionic DM is FIMP and pseudo scalar particle $\rho$ enters thermal equilibrium so act as a WIMP.

The main difference between freeze-out and freeze-in scenario is the DM particles does not reach equilibrium in freeze-in. This means that for FIMP particles as DM, we have $Y\ll Y_{eq}$. In \cite{Hall:2009bx}, it has been shown that abundance of FIMP DM particles satisfies the following condition: 
\begin{eqnarray}
Y_{FI}\sim\lambda^2(\frac{M_{pl}}{m})
\label{FC}
\end{eqnarray}
Therefore the very feeble coupling of the FIMP is generic feature of freeze-in mechanism. In other word, to get the sense of viability of freeze-in condition, we should note that decay time is of the same  order to  Hubble time. Considering $T_e\sim Y_{FI}m$ where $T_e$ is  the temperature for matter- radiation equality, the coupling condition for our model leads to:
\begin{eqnarray}
g_s\lambda_1\sim\sqrt{\frac{T_e}{M_{pl}}}(\frac{m}{v_H}).
\label{Con}
\end{eqnarray}
This condition has been considered for the rest of analysis.

In \cite{Ghorbani:2014qpa}, stability and perturbativty conditions on parameters space have been studied. We consider the $m_{\chi}$, $m_{\rho}$ (mass parameters), $g_p$, $\lambda_1$ and $\lambda$ (coupling parameters) as parameters of the model and set their values to satisfy stability and perturbativity conditions ($\lambda\lambda_H>6\lambda^2_1$ and $|\lambda_i|<4\pi$).
 In addition, we display the lower bound on the mass of DM particles, produced from various dwarf spheroidal galaxies \cite{Lower bound DM}. In this scenario, we suppose that $m_{DM}>1.7~\rm KeV$.
\subsection{Non-thermal production of the pseudo-scalar}
In this case, we suppose that $g_p$ and $\lambda_1$ are very small. This means $\chi$ and $\varphi$ interact feebly and the dominant contribution to the relic density of DM is generated via freeze-in mechanism. The pseudo scalars are produced non-thermally which decay into DM. Therefore, we can neglect initial abundance of pseudo scalar particles in Boltzmann equation for $\chi$. We can write the Boltzmann equation for $\chi$ and $\rho$ as:
\begin{eqnarray}
\frac{d n_{\chi}}{dt}+3Hn_{\chi}&=&\frac{m^2_hT}{\pi^2}K_1(\frac{m_h}{T})\Gamma_{h\rightarrow \chi\chi}+\sum_{j=Z,W,f,h}\frac{T}{32\pi^4}\nonumber\\&&\int^\infty_{4m^2_{j}}ds\sigma_{jj\rightarrow \chi\chi}(s)(s-4m^2_{j})\sqrt{s}K_1(\frac{\sqrt{s}}{T}),
\label{Boltzman-nonth}
\end{eqnarray}
\begin{eqnarray}
\frac{d n_{\rho}}{dt}+3Hn_{\rho}&=&\frac{m^2_hT}{\pi^2}K_1(\frac{m_h}{T})\Gamma_{h\rightarrow \rho\rho}-\frac{m^2_{\rho}T}{\pi^2}K_1(\frac{m_{\rho}}{T})\Gamma_{\rho\rightarrow \chi\chi}+\sum_{j=Z,W,f,h}\frac{T}{32\pi^4}\nonumber\\&&\int^\infty_{4m^2_{j}}ds\sigma_{jj\rightarrow \rho\rho}(s)(s-4m^2_{j})\sqrt{s}K_1(\frac{\sqrt{s}}{T})
\label{Boltzman-r},
\end{eqnarray}
where $K_1$ is the modified Bsssel function of order 1 and $\sqrt{s}$ is center of mass energy. As it is seen, above equations are not coupled because of earlier assumption about couplings. In the following, we discuss evolution of $n_{\rho}$ and $n_{\chi}$. The decay widths $\Gamma_{i\rightarrow jj}$ and annihilation cross sections $\sigma_{ii\rightarrow jj}$ are given in Appendix.
We can rewrite the Boltzmann equation for fermionic DM in terms of yield, $Y=n/s$:
\begin{eqnarray}
Y_{\chi}&=&\frac{1}{4\pi^4}\frac{45M_{pl}}{1.66g^{s}_*(T)\sqrt{g^{\rho}_*}}[2\Gamma_{h\rightarrow \chi\chi}m^2_h\int^\infty_{T_{Now}}dT \frac{K_1(\frac{m_h}{T})}{T^5}+\sum_{j=Z,W,f,h}\frac{1}{16\pi^2}\nonumber\\&&\int^\infty_{T_{Now}}dT \frac{1}{T^5} \int^\infty_{4m^2_i}ds\sigma_{jj\rightarrow \chi\chi}(s)(s-4m^2_{i})\sqrt{s}K_1(\frac{\sqrt{s}}{T})],
\label{Boltzman-k}
\end{eqnarray}
where $M_{Pl}$ is the Plank mass and $g^s_*$ and $g^{\rho}_*$ are the effective numbers for degrees of freedom and using
$\dot{T}=-HT$, $H=1.66\sqrt{g^{\rho}_*}\frac{T^2}{M_{pl}}$ and $S=g^{s}_*(T)\frac{2\pi^2}{45}T^3$.

\begin{figure}
\begin{center}
\centerline{\hspace{0cm}\epsfig{figure=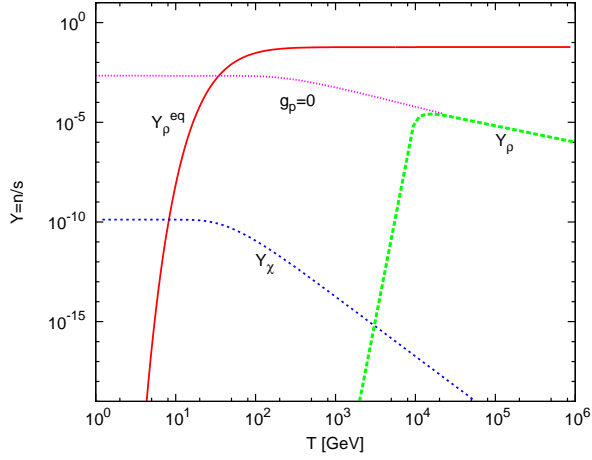,width=8cm}}
\end{center}
\caption{$\rho$ and $\chi$ abundance as a function of temperature for non-thermal production of pseudo scalar particles $\rho$. In this figure, $m_{DM}=20~\rm GeV$, $m_{\rho}=200~\rm GeV$, $\lambda=5.3$, $\lambda_1=10^{-6}$ and  $g_p=3\times10^{-7}$.}\label{YrhoFF}
\end{figure}

In Fig.~\ref{YrhoFF}, we display yield quantity as a function of temperature for  non-thermally produce  $\chi$ and $\rho$. As the time passes and Universe cools down, DM abundance steadily increases. By the time freeze-in starts, $\rho$ fails to maintain its freeze-in density as it decays to $\rho\rightarrow\chi\chi$. As it is expected, the freeze-in temperature is approximately equal to three times of DM mass and is independent of couplings. For $g_p=0$ and $v_\varphi=0$, there is not mixing between Higgs field and pseudo scalar fields and as a result, $\rho$ is stable and solely plays role of DM and after freeze-in, $\rho$ abundance remains constant. Note that in this case, $\rho$ interacts with SM Higgs but can not decay to other SM particles.  

\begin{figure}
\begin{center}
\centerline{\hspace{0cm}\epsfig{figure=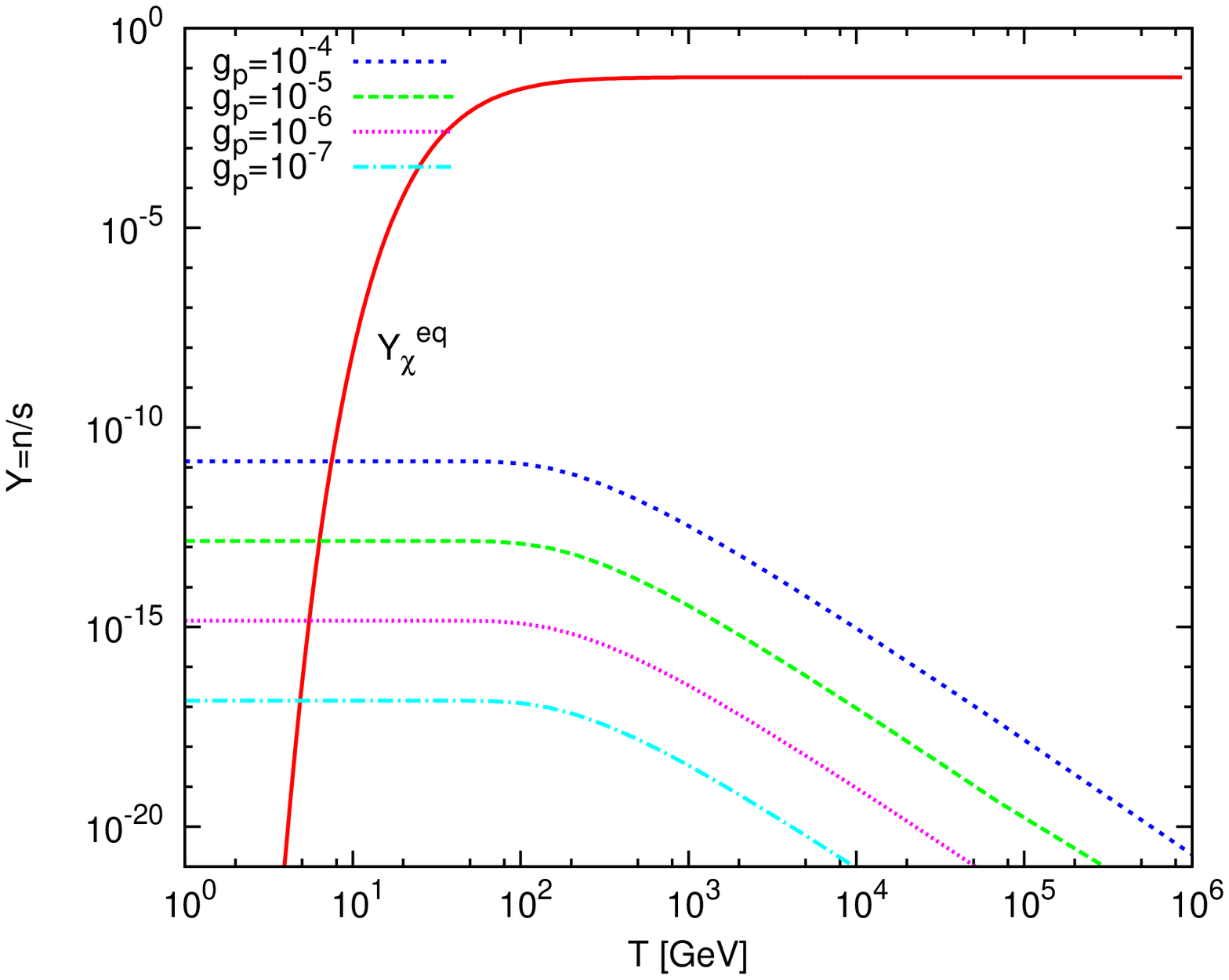,width=6.5cm}\hspace{0cm}\epsfig{figure=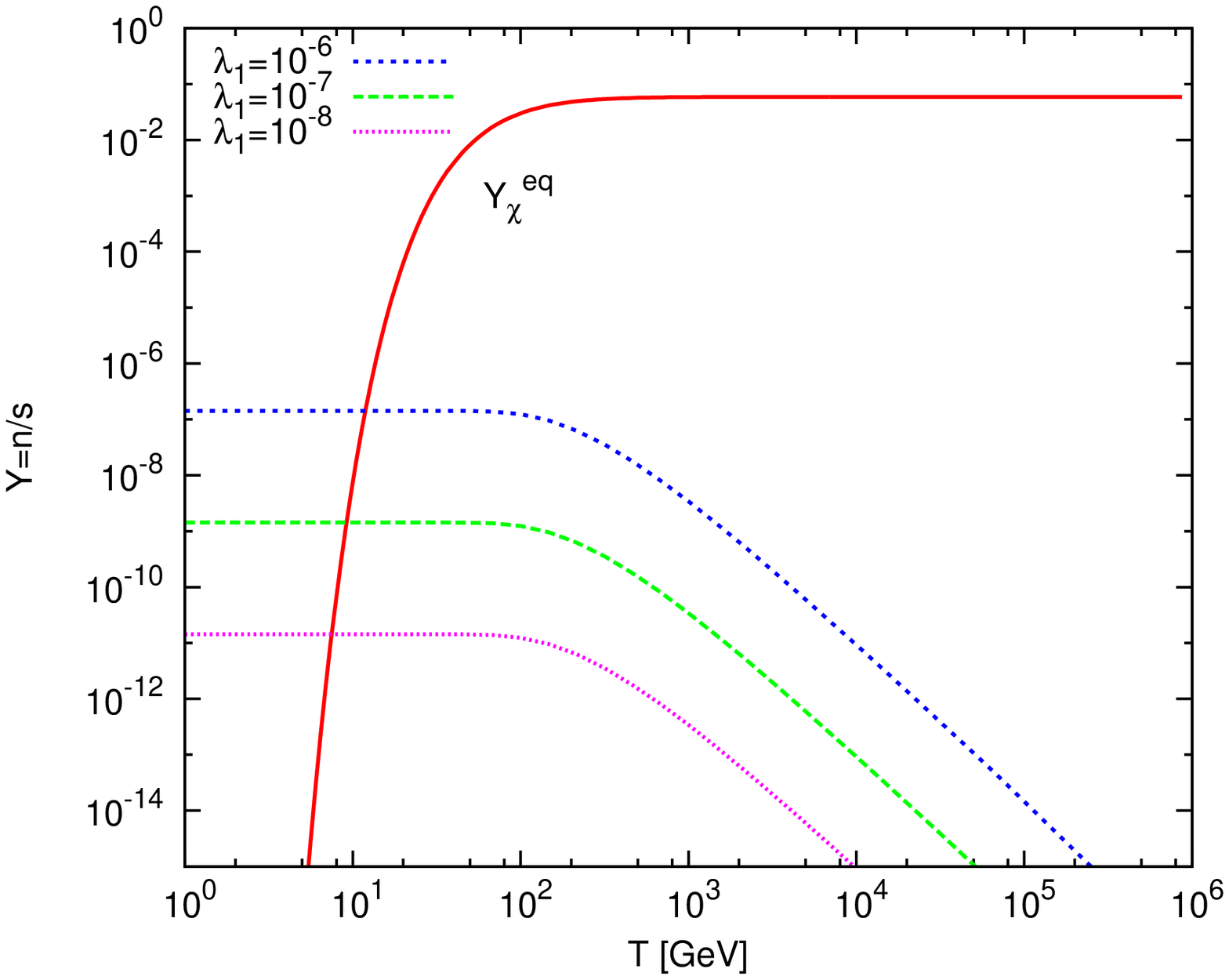,width=6.5cm}}
\centerline{\vspace{-1.5cm}\hspace{0.5cm}(a)\hspace{6cm}(b)}
\centerline{\vspace{-0.0cm}}
\end{center}
\caption{The $\chi$ abundance as a function of temperature for non-thermal production of pseudo scalar particles $\rho$. We set for all figures $m_{\rho}=150~\rm GeV$, $\lambda=3$ and $m_{DM}=200~\rm GeV$. a) for $\lambda_1=3\times10^{-7}$ and different values of $g_p$. b) for $g_p=10^{-3}$ and different values of mixing couplings $\lambda_1$.}\label{Yc}
\end{figure}

In Fig.~\ref{Yc}, we display the DM abundance as a function of $T$ for $m_{DM}=200~\rm GeV$ and $\lambda=3$. As it is seen, in all plots, condition $Y_{DM}\ll Y^{eq}_{DM}$ (freeze-in condition) is satisfied.

\begin{figure}
\begin{center}
\centerline{\hspace{0cm}\epsfig{figure=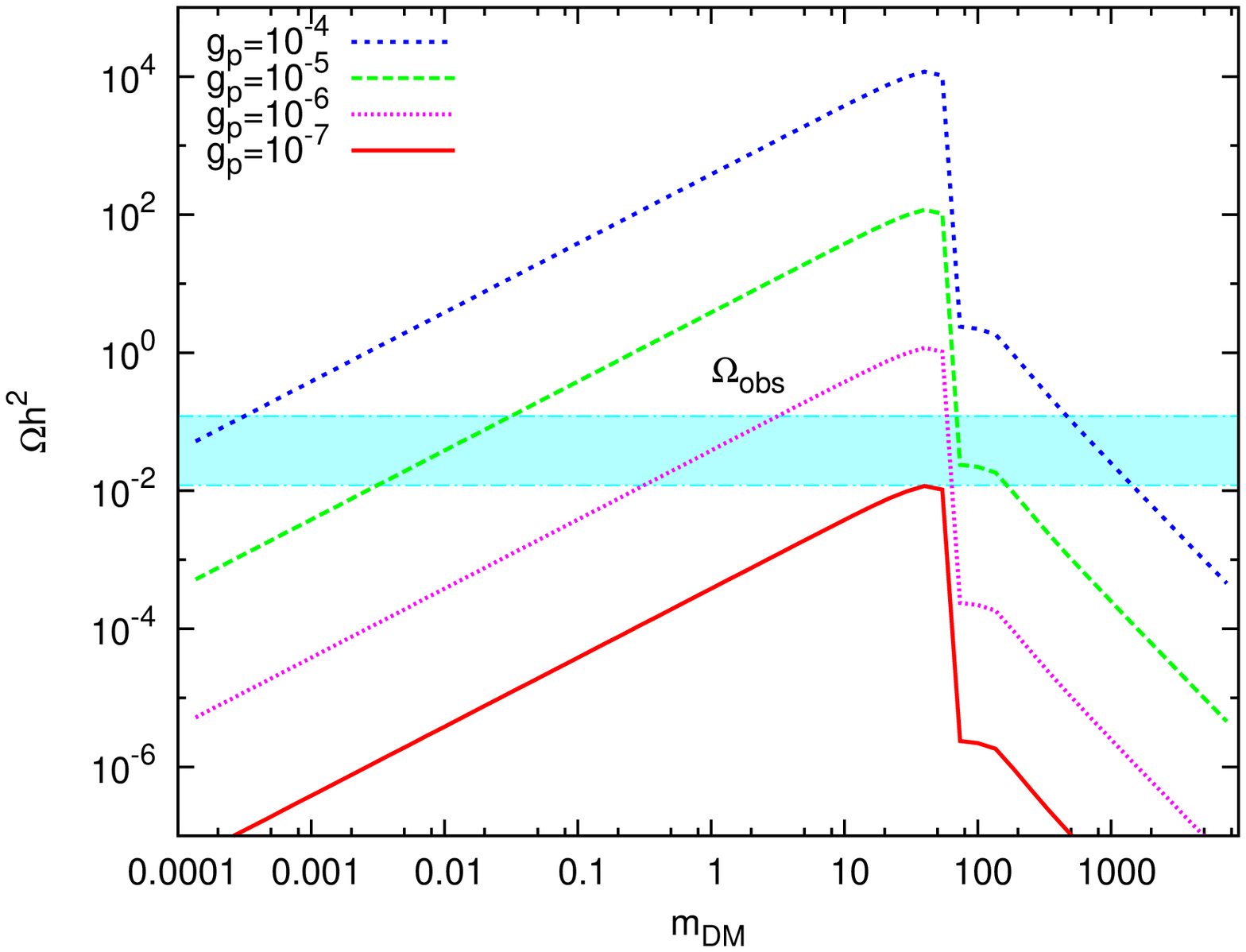,width=6.5cm}\hspace{0cm}\epsfig{figure=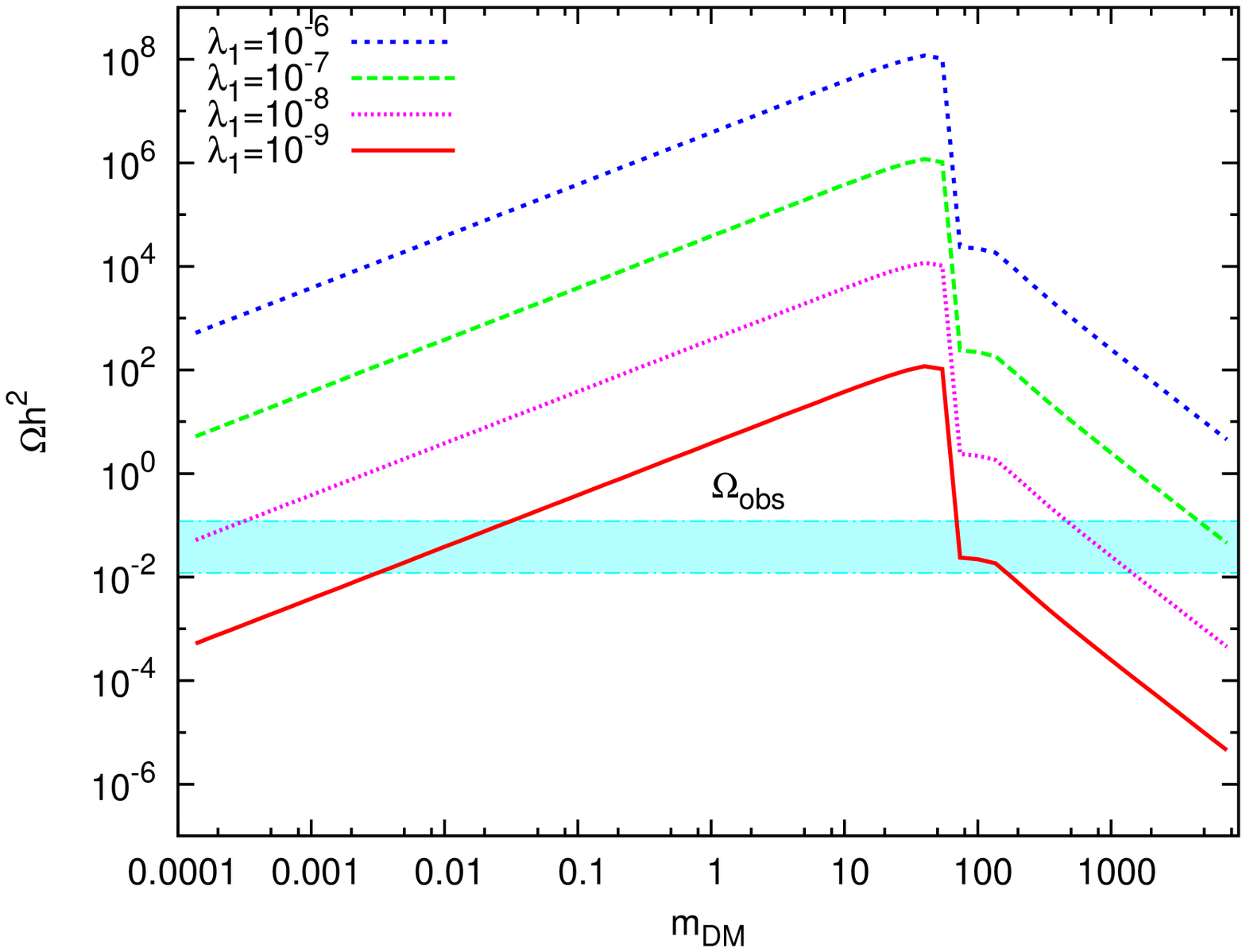,width=6.5cm}}
\centerline{\vspace{-1.5cm}\hspace{0.5cm}(a)\hspace{6cm}(b)}
\centerline{\vspace{-0.0cm}}
\end{center}
\caption{The relic density as a function of $m_{\chi}$ for non-thermal production of pseudo scalar particles $\rho$. The shadowed panel indicates regions in which $\chi$ particles contribute more than 10 percent of dark matter density. In this figure, $m_{\rho}=150~\rm GeV$ and $\lambda=3$. a) for $\lambda_1=10^{-7}$ and different values of $g_p$. b) for $g_p=10^{-3}$ and different values of mixing coupling $\lambda_1$. }\label{Omega-non}
\end{figure}

The relic density of DM is well measured by WMAP and Planck experiments and the current value is $\Omega_{DM}h^2 = 0.1199 \pm 0.0027$ \cite{PLANCK},
where $h = 0.67 \pm 0.012$ is the scaled current Hubble parameter in units of $100\rm km/s.Mpc$.
The current relic density of DM can be obtained from:
\begin{equation}
\Omega_{DM}h^2 =2.742\times10^{-8}(\frac{M_{\chi_i}}{GeV}) Y(T_0).
\label{Re}
\end{equation}
Fig.~\ref{Omega-non} indicates how our model contributes to DM density for the different values of couplings. For region below the resonance, relic density increases linearly with DM mass and for above resonance, relic density decreases with DM mass.  For $m_\chi<m_h/2$ the $h\rightarrow\chi\chi$ is the main production source for the DM. This means that  below the resonance, relic density is proportional to DM mass and increasing the DM mass would lead to larger abundance.

From the figure, it can be seen that after the resonance region $h\rightarrow\chi\chi$ is kinematically  suppressed which causes exponentially the declination of relic abundance. 
In our analysis, we use Maxwell-Boltzmann distribution which is suitable to cold DM as non relativistic particles. Due to numerical complication in analysis of  relativistic particles with F-D distribution, we avoided study of light mass DM (KeV) in Fig.~\ref{Omega-non}-b (For detail analysis see \cite{merle}).

\subsection{Thermal production of the pseudo-scalar}

In this case, we suppose that the mixing coupling between Higgs scalar and pseudo scalar field is strong enough to produce $\rho$ thermally ($\lambda_1\geq10^{-3}$). The large coupling implies that $f_\rho=f^{eq}_\rho$, and initial number density would not influence our results. The Boltzmann equations can be written as:
\begin{eqnarray}
\frac{d n_{\chi}}{dt}+3Hn_{\chi}&=&\Gamma_{\rho\rightarrow \chi\chi} n_{\rho,eq}+\Gamma_{h\rightarrow \chi\chi} n_{h,eq}+\sum_{j=Z,W,f,h} \langle\sigma_{jj\rightarrow\chi\chi}\upsilon\rangle \nonumber\\&&n^2_{j,eq}+ \langle\sigma_{\rho\rho\rightarrow\chi\chi}\upsilon\rangle n^2_{\rho,eq}+\langle\sigma_{h\rho\rightarrow\chi\chi}\upsilon\rangle n_{h,eq}n_{\rho,eq},
\label{Boltzman-k-roeq}
\end{eqnarray}
\begin{eqnarray}
\frac{d n_{\rho}}{dt}+3Hn_{\rho}&=&-\Gamma_{\rho\rightarrow \chi\chi} n_{\rho}-\Gamma_{\rho\rightarrow hh}(n_{\rho}-n^{eq}_{\rho})\nonumber\\&-&\sum_{j=Z,W,f,h} \langle\sigma_{\rho\rho\rightarrow jj}\upsilon\rangle (n^2_{\rho}-n^2_{\rho,eq})-\langle\sigma_{\rho\rho\rightarrow \chi\chi}\upsilon\rangle n^2_{\rho}\nonumber\\&-&\langle\sigma_{h\rho\rightarrow \chi\chi}\upsilon\rangle n_{\rho}n_{h,eq}
\label{Boltzman-r},
\end{eqnarray}
The underlying assumption in this case, would help us avoid complications to solve two coupled Boltzman equations. We can rewrite Boltzmann equation for $\chi$ as:
\begin{eqnarray}
&&\frac{d n_{\chi}}{dt}+3Hn_{\chi}=\sum_{i=h,\rho}\frac{m^2_i T}{\pi^2}K_1(\frac{m_h}{T})\Gamma_{h\rightarrow \chi\chi}+\sum_{j=Z,W,f,h,\rho}\frac{T}{32\pi^4}\nonumber\\&&\int^\infty_{4m^2_{i}}ds\sigma_{jj\rightarrow \chi\chi}(s)(s-4m^2_{i})\sqrt{s}K_1(\frac{\sqrt{s}}{T})+\frac{T}{32\pi^4}\nonumber\\&&\int^\infty_{(m_{\rho}+m_{h})^2}ds\sigma_{h\rho\rightarrow \chi\chi}(s)(s-(m_{\rho}+m_{h})^2)\sqrt{s}K_1(\frac{\sqrt{s}}{T}).
\label{Boltzman-nonth}
\end{eqnarray}
Rewriting the Boltzmann equation in terms of yield, we have:
\begin{eqnarray}
&&Y_{\chi}=\frac{1}{4\pi^4}\frac{45M_{pl}}{1.66g^{s}_*(T)\sqrt{g^{\rho}_*}}[2\Gamma_{h\rightarrow \chi\chi}m^2_h\int^\infty_{T_{Now}}dT \frac{K_1(\frac{m_h}{T})}{T^5}+\sum_{j=Z,W,f,h}\frac{1}{16\pi^2}\nonumber\\&&\int^\infty_{T_{Now}}dT \frac{1}{T^5} \int^\infty_{4m^2_i}ds\sigma_{jj\rightarrow \chi\chi}(s)(s-4m^2_{i})\sqrt{s}K_1(\frac{\sqrt{s}}{T})]+\frac{1}{16\pi^2}\nonumber\\&&\int^\infty_{T_{Now}}dT \frac{1}{T^5} \int^\infty_{(m_{\rho}+m_{h})^2}ds\sigma_{h\rho\rightarrow \chi\chi}(s)(s-(m_{\rho}+m_{h})^2))\sqrt{s}K_1(\frac{\sqrt{s}}{T})].
\label{Y-nonth}
\end{eqnarray}

\begin{figure}
\begin{center}
\centerline{\hspace{0cm}\epsfig{figure=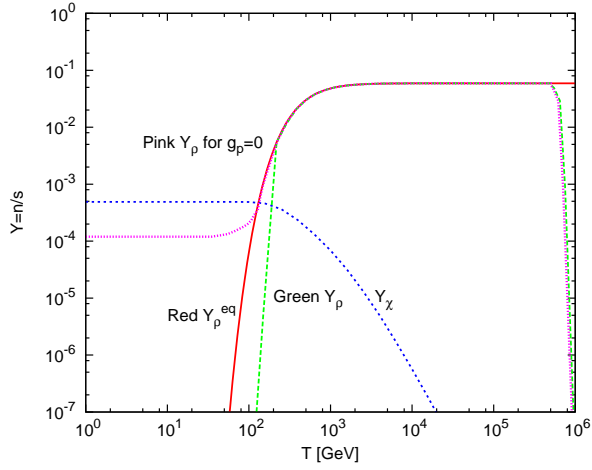,width=8cm}}
\end{center}
\caption{$\rho$ and $\chi$ abundance as a function of temperature for thermal production of pseudo scalar particles $\rho$. In this figure, we set $\lambda=3$, $\lambda_1=0.13$, $g_p=5\times10^{-9}$, $m_{DM}=200~\rm GeV$ and  $m_{\rho}=1000~\rm GeV$.}\label{YrhoFW}
\end{figure}
In Fig.~\ref{YrhoFW}, we have shown yield quantity as a function of temperature in case that pseudo scalar produce thermally  for $\chi$ and $\rho$.
As it is shown in figure, for $g_p=0$ and $v_{\varphi}=0$, $\rho$ is the only existing DM which freeze-out like a WIPM.

\begin{figure}
\begin{center}
\centerline{\hspace{0cm}\epsfig{figure=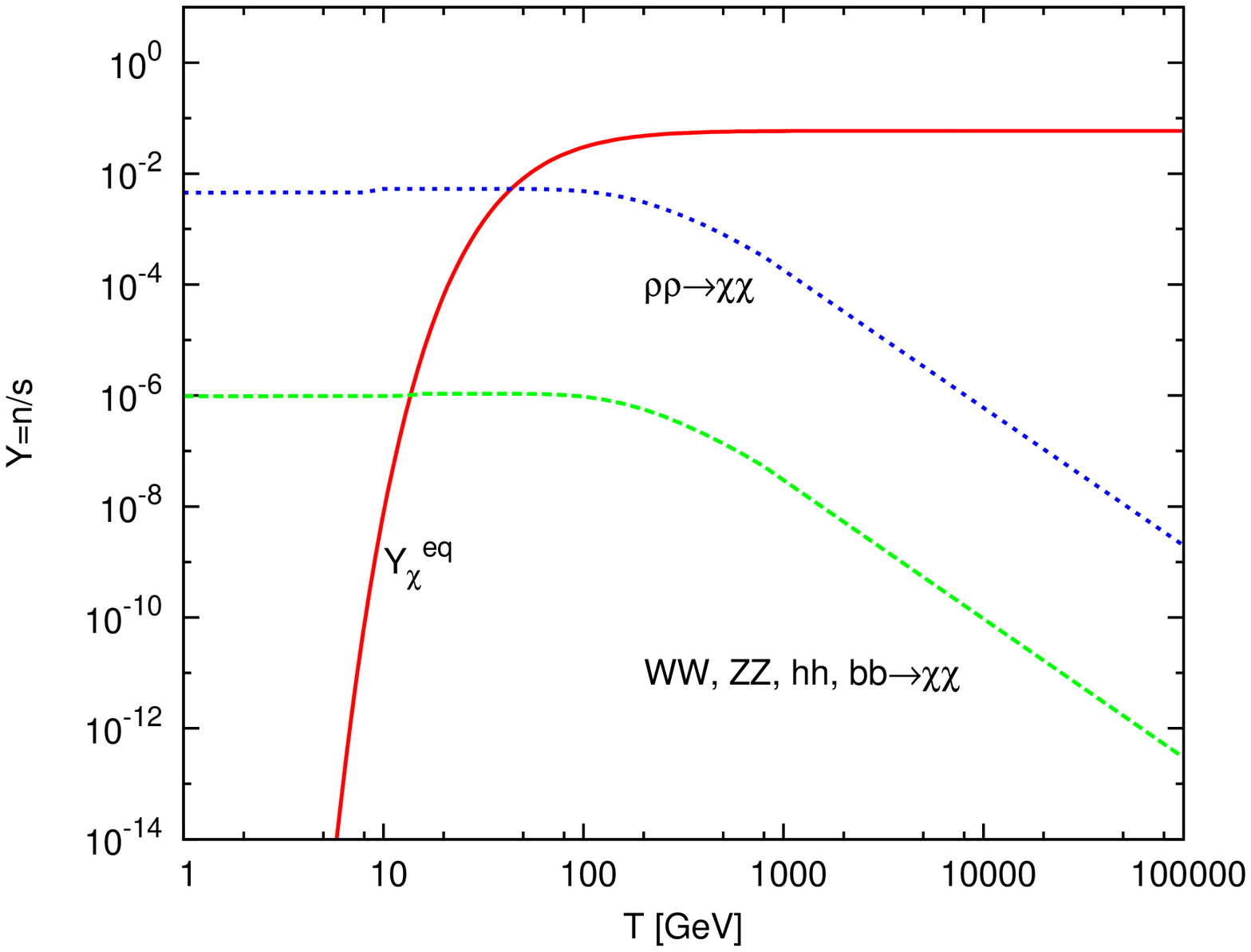,width=6.5cm}\hspace{0cm}\epsfig{figure=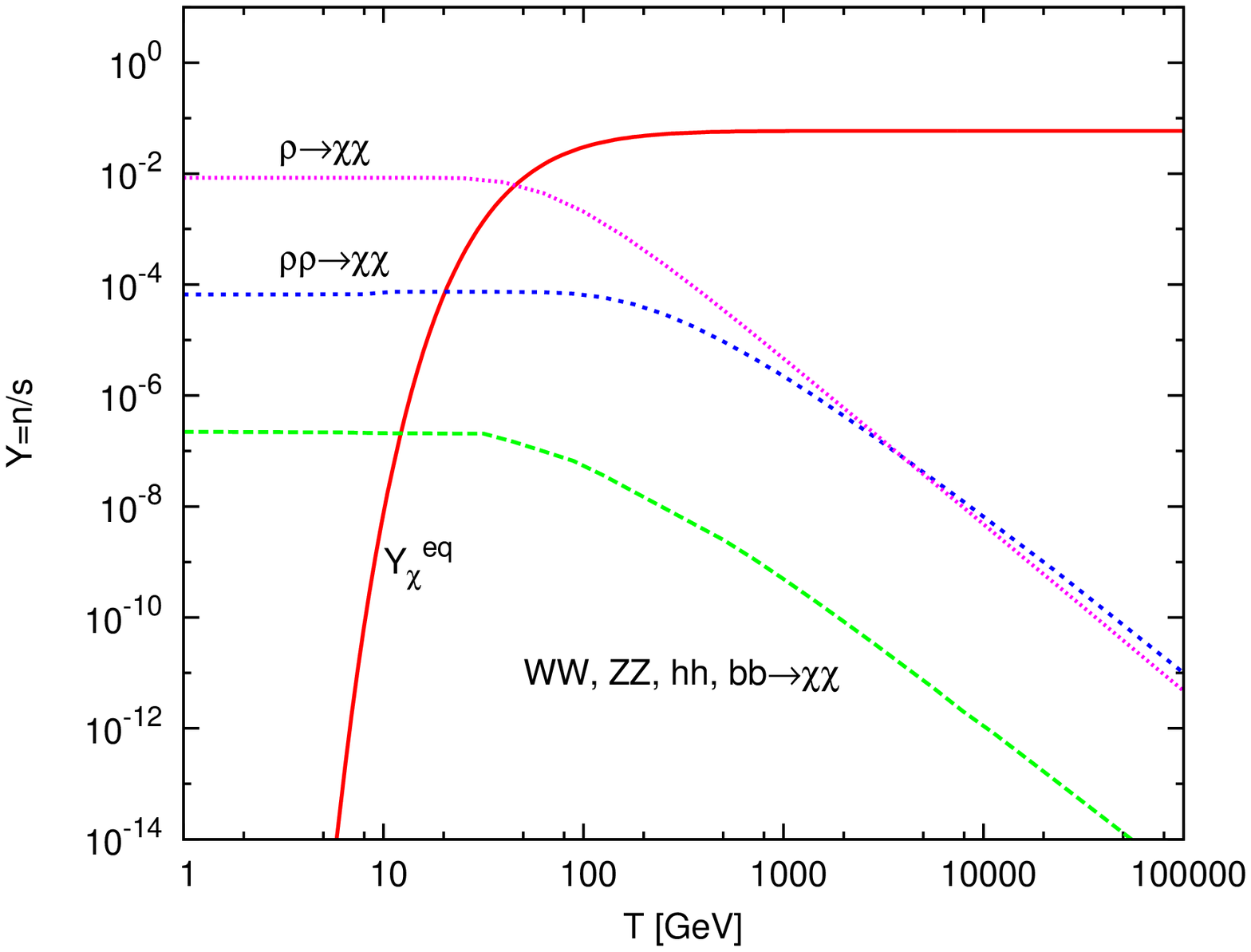,width=6.5cm}}
\centerline{\vspace{-1.5cm}\hspace{0.5cm}(a)\hspace{6cm}(b)}
\centerline{\vspace{-0.0cm}}
\end{center}
\caption{The $\chi$ abundance as a function of temperature for thermal production of pseudo scalar particles $\rho$. Input parameters are: $m_{\rho}=200~\rm GeV$, $g_p=10^{-7}$, $\lambda_1=0.003$ and $\lambda=3$ a) for $m_{DM}=200~\rm GeV$. b) for $m_{DM}=20~\rm GeV$ which $\rho\rightarrow \chi\chi$ occurs. }\label{DC}
\end{figure}

In the case in which $\rho$ is produced thermally, different processes may contribute to the DM production. Fig.~\ref{DC} represents relevant contribution of different channel in the yields quantity. In Fig.~\ref{DC}-a, since $\rho$ cannot decay to DM, the main contribution to DM abundance arises from annihilation of $\rho\rho$ and other processes such as annihilation of $WW$, $ZZ$, $hh$ or $bb$ are subdominant.

\begin{figure}
\begin{center}
\centerline{\hspace{0cm}\epsfig{figure=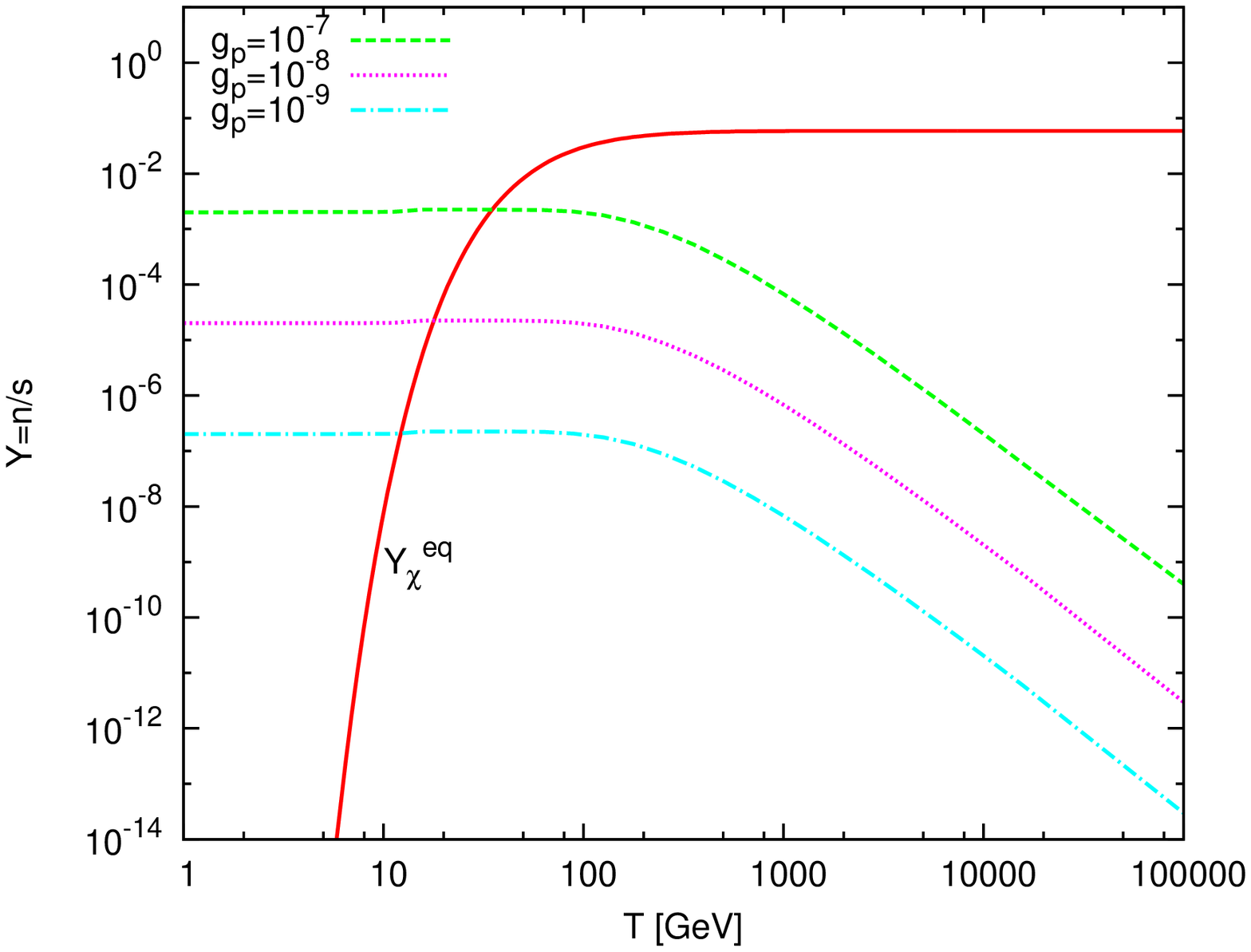,width=6.5cm}\hspace{0cm}\epsfig{figure=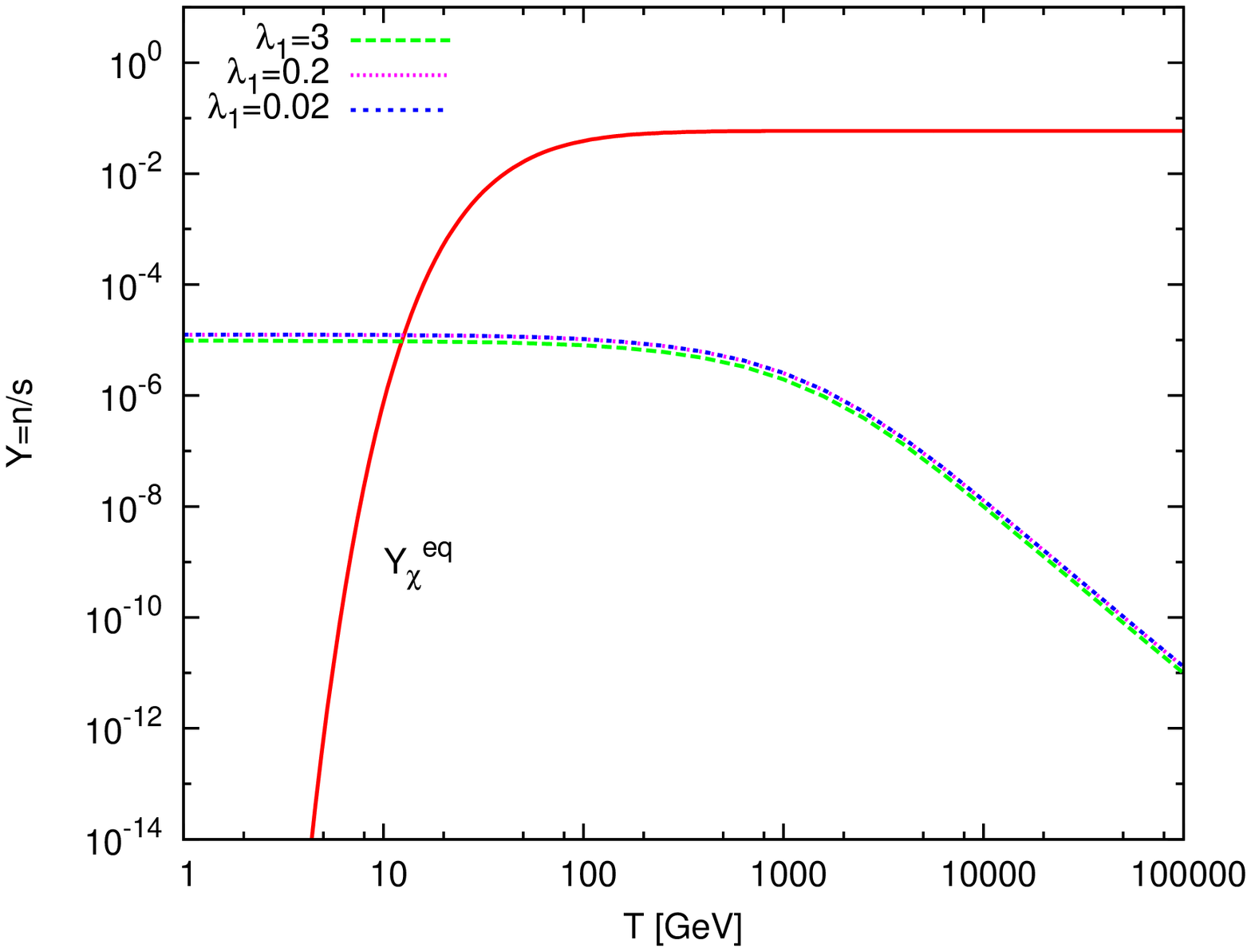,width=6.5cm}}
\centerline{\vspace{-1.5cm}\hspace{0.5cm}(a)\hspace{6cm}(b)}
\centerline{\vspace{-0.0cm}}
\end{center}
\caption{The $\chi$ abundance as a function of temperature for thermal production of pseudo scalar particles $\rho$. Input parameters are: $m_{DM}=200~\rm GeV$, $m_{\rho}=150~\rm GeV$ and $\lambda=3$ a) for $\lambda_1=0.001$ and different values of $g_p$ b) for $m_{DM}=150~\rm GeV$, $m_{\rho}=400~\rm GeV$, $\lambda=20$ and $g_p=10^{-8}$ and different values of mixing coupling $\lambda_1$.) }\label{Y-thermal}
\end{figure}

Fig.~\ref{Y-thermal} shows yield quantity as a function of temperature. As it is seen in Fig.~\ref{DC}, for $m_{\rho}>2m_{\chi}$ the main contribution to DM abundance arises from $\rho\rightarrow\chi\chi$ which is proportional to $\rm cos~\theta$. For small value of $\theta$ ($\lambda_1\propto2\theta$), dependency of this function  to $\theta$ is very small which is observable in Fig.~\ref{Y-thermal}-b.
\begin{figure}
\begin{center}
\centerline{\hspace{0cm}\epsfig{figure=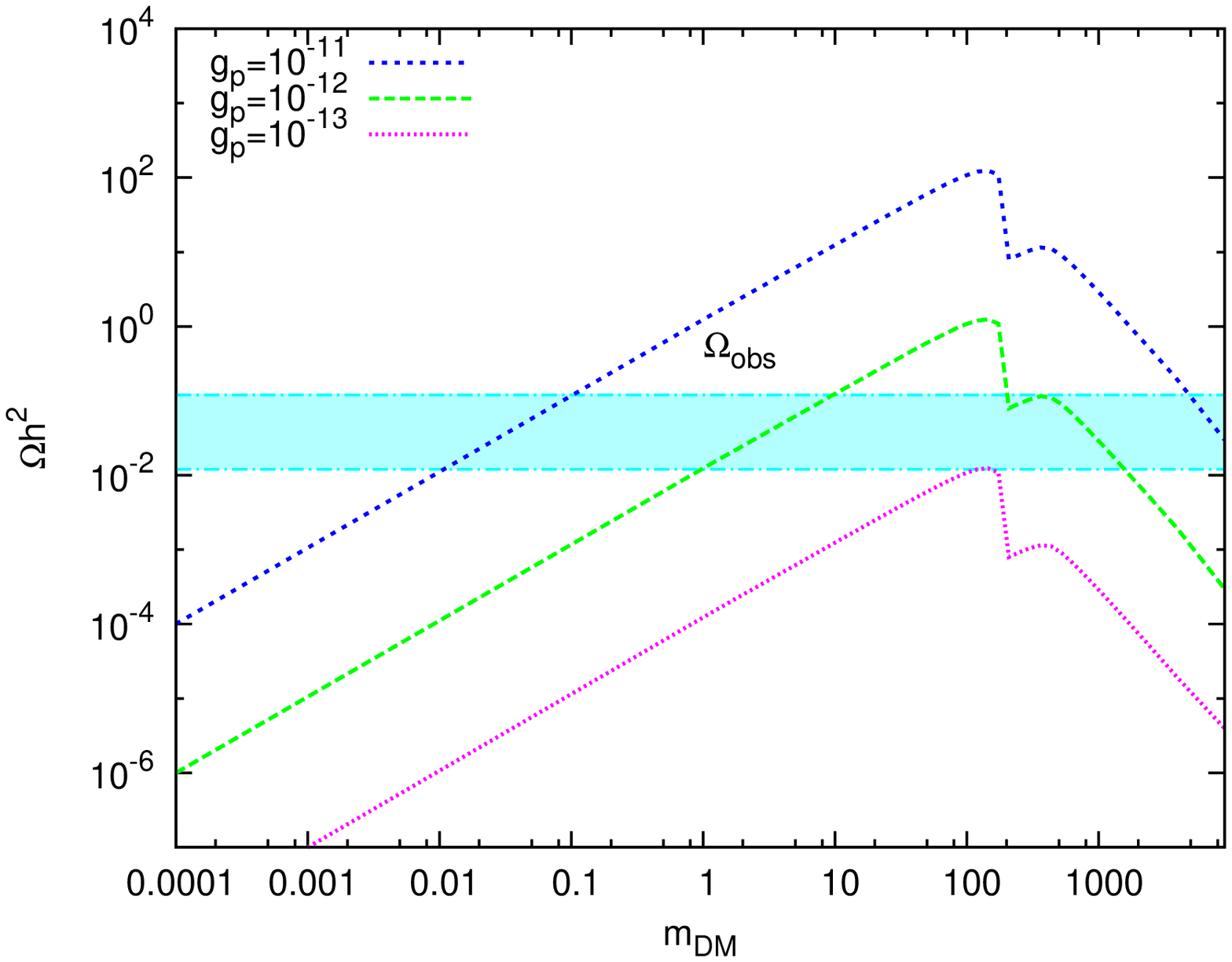,width=6.5cm}\hspace{0cm}\epsfig{figure=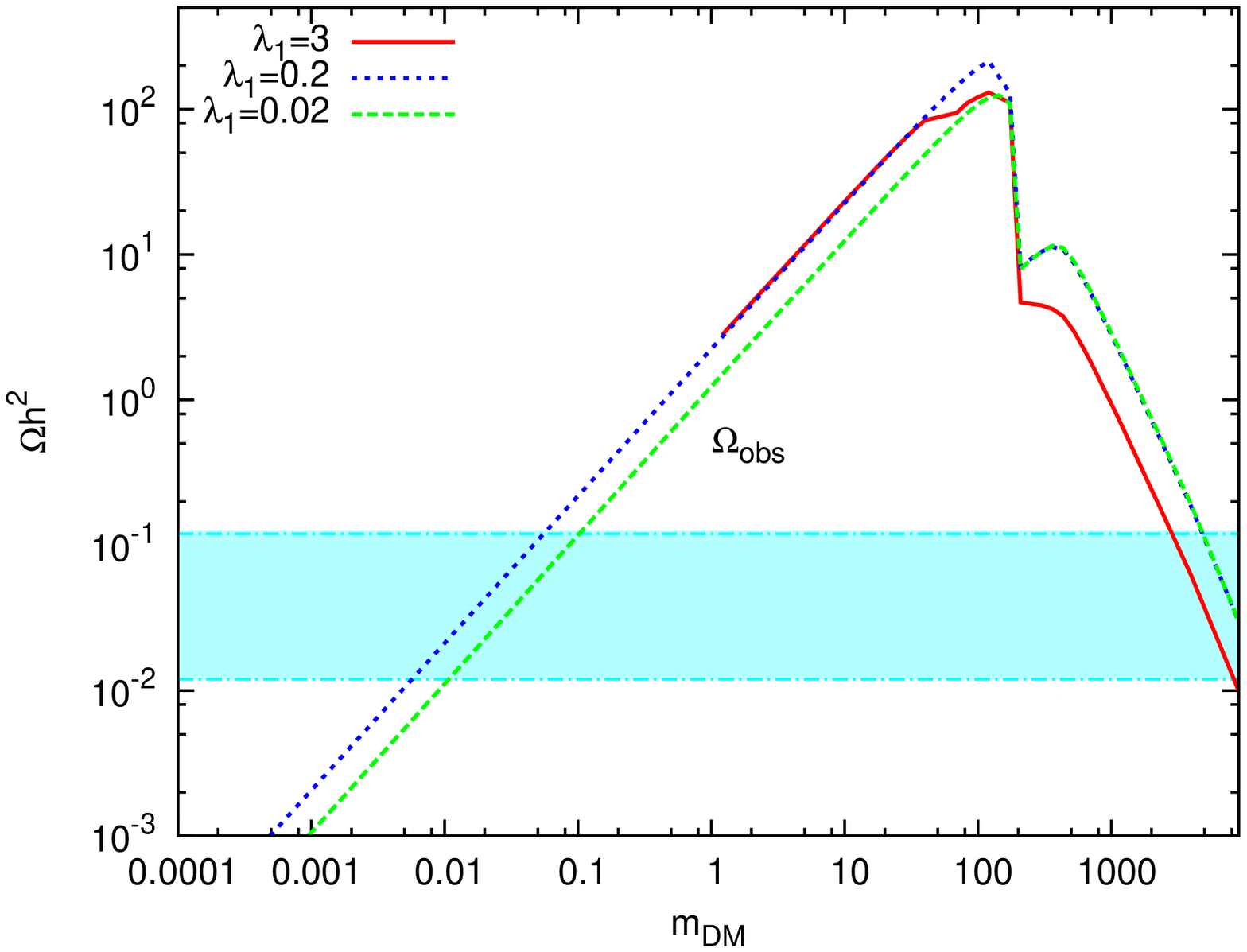,width=6.5cm}}
\centerline{\vspace{-1.5cm}\hspace{0.5cm}(a)\hspace{6cm}(b)}
\centerline{\vspace{-0.0cm}}
\end{center}
\caption{The relic density as a function of $m_{\chi}$ for thermal production of pseudo scalar particles $\rho$. The shadowed panel indicates regions in which $\chi$ particles contribute more than 10 percent of dark matter density. In this figure $m_{\rho}=400~\rm GeV$ and $\lambda=3$, a) for $\lambda_1=0.2$ and different values of $g_p$. b) for $g_p=10^{-11}$ and different values of mixing coupling $\lambda_1$. }\label{Omega-therm}
\end{figure}
Fig.~\ref{Omega-therm} depicts DM relic density for different values of couplings.
\section{Phenomenological aspects}
\subsection{Direct detection}
In our model, DM can interact with nucleon by Higgs boson or the pseudo scalar exchange. The expression for spin independent elastic cross section between fermionic DM and nucleon N is given by \cite{Ghorbani:2014qpa}:
\begin{equation}
\sigma_{SI} = \frac{g_p^2 \sin^2\theta\cos^2 \theta f(n)^2}{4\pi}\frac{m^2_N}{v^2_H}(\frac{1}{m^2_h}-\frac{1}{m^2_{\rho}})^2\frac{2v_{rel}^2\mu^4}{m^2_{\chi}},
\label{si}
\end{equation}
where the coupling constant $f(n)$ is given by nuclear matrix elements \cite{Higgs-nucleon}, $v$ is DM-nucleus relative velocity  and $\mu=m_Nm_{\chi}/(m_N+m_{\chi})$ is reduced mass of DM-nucleon. As it is mentioned, for freeze-in mechanism the coupling of DM to pseudo scalar fields is extremely small. Since the DM-nucleon cross section is proportional with $g^2_p\sin^2\theta$ and relative velocity ($\upsilon_{rel}\sim {\cal O}(10^{-3})$), direct cross section of FIMP-DM  is smaller by a factor $ {\cal O}(10^{-16})$ than usual sensitivity of current experiments such as $\rm XENON100$ \cite{XENON100} and $\rm LUX$ \cite{LUX}. Therefore fermionic DM candidate in our model is not detectable by direct detection. This is a desirable outcome since the direct detection experiment has not reported any signal so far.

\begin{figure}
\begin{center}
\centerline{\hspace{0cm}\epsfig{figure=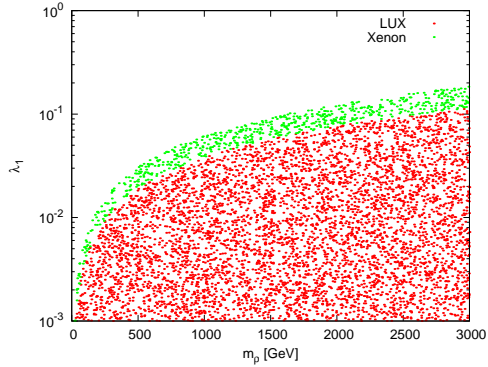,width=6.5cm}}
\centerline{\vspace{-1.5cm}}
\end{center}
\caption{Scater points depict ranges of parameters space in mass of pseudo-scalar and  mixing coupling $|\lambda_1|$  plane which are consistent with $\rm XENON100$ \cite{XENON100} and $\rm LUX$ \cite{LUX} experiment.}\label{ScaterDirect}
\end{figure}

As it was mentioned, in the case that pseudo scalar produce thermally  and  $g_p=0$ and $v_{\varphi}=0$, $\rho$ can play role of DM as WIPM and therefore it can be detectable in direct search experiments.  This feature have been studied in Fig.~\ref{ScaterDirect}.

\subsection{Invisible Higgs decay}
The studying of SM Higgs boson properties  is being pursued as a main window for new physics searches. If the mass of the new particle is less than half of the Higgs mass, the Higgs could decay into the light DM pairs with a large invisible branching ratio. In our model, for $m_{\rho}$ or $m_{\chi}<m_h/2$, they can contribute to the invisible decay mode of Higgs boson with branching ratio:
\begin{eqnarray}
Br(h\rightarrow \rm Invisible)& =\frac{\Gamma(h\rightarrow \chi\chi)+\Gamma(h(\rightarrow \rho\rho)}{\Gamma(h)_{SM}+\Gamma(h\rightarrow \chi\chi)+\Gamma(h\rightarrow \rho\rho)},
\label{decayinv1}
\end{eqnarray}
where $\Gamma(h)_{SM}=4.15 ~ \rm [MeV]$ is total width of Higgs boson \cite{SM Higgs branching ratio} and partial width for processes $h\rightarrow \rho\rho$ and $h\rightarrow \chi\chi$  are given by:
\begin{eqnarray}
\Gamma(h\rightarrow \chi\chi)=\frac{m_h g_p^2\sin ^2\theta }{8\pi }(1-4m_\chi^2/m_h^2)^\frac{1}{2},
\end{eqnarray}
\begin{eqnarray}
\Gamma(h\rightarrow \rho\rho)=\frac{c^2}{16\pi m_h}(1-4m_\rho^2/m_h^2)^\frac{1}{2}.
\end{eqnarray}

The SM prediction for Higgs branching ratio to invisible particles originates from process $h\rightarrow ZZ^*\rightarrow 4\nu$  is \cite{Higgs branching ratio}:
\begin{eqnarray}
Br(h\rightarrow ZZ^*\rightarrow 4\nu)=1.2\times10^{-3}.
\label{amp1}
\end{eqnarray}
This quantity has been constrained by various groups using the latest data from LHC \cite{CMS:2015dia}-\cite{Invisible Higgs decay Exp}.  The ATLAS Collaboration results lead to an upper limit of $0.29$ at $95\%$ C.L \cite{Invisible Higgs decay Exp}. In our scenario, coupling of DM with Higgs boson is very small, so the contribution of DM to invisible Higgs decay is negligible. Nevertheless in the scenario which pseudo-scalar particle is produced thermally, mixing coupling with Higgs boson is large and $h\rightarrow \rho\rho$ can contribute to invisible Higgs decay.

\begin{figure}
\begin{center}
\centerline{\hspace{0cm}\epsfig{figure=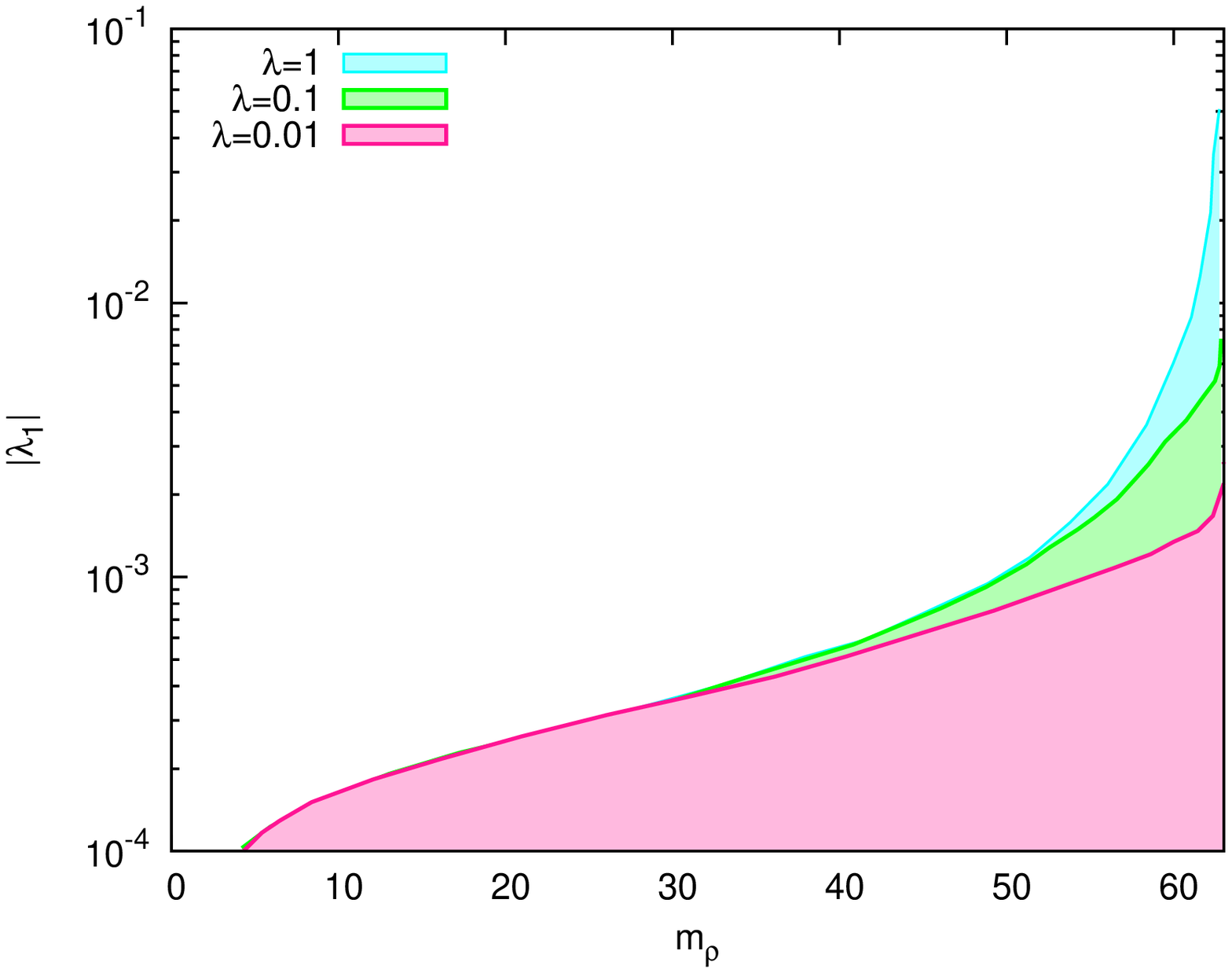,width=6.5cm}\hspace{0cm}\epsfig{figure=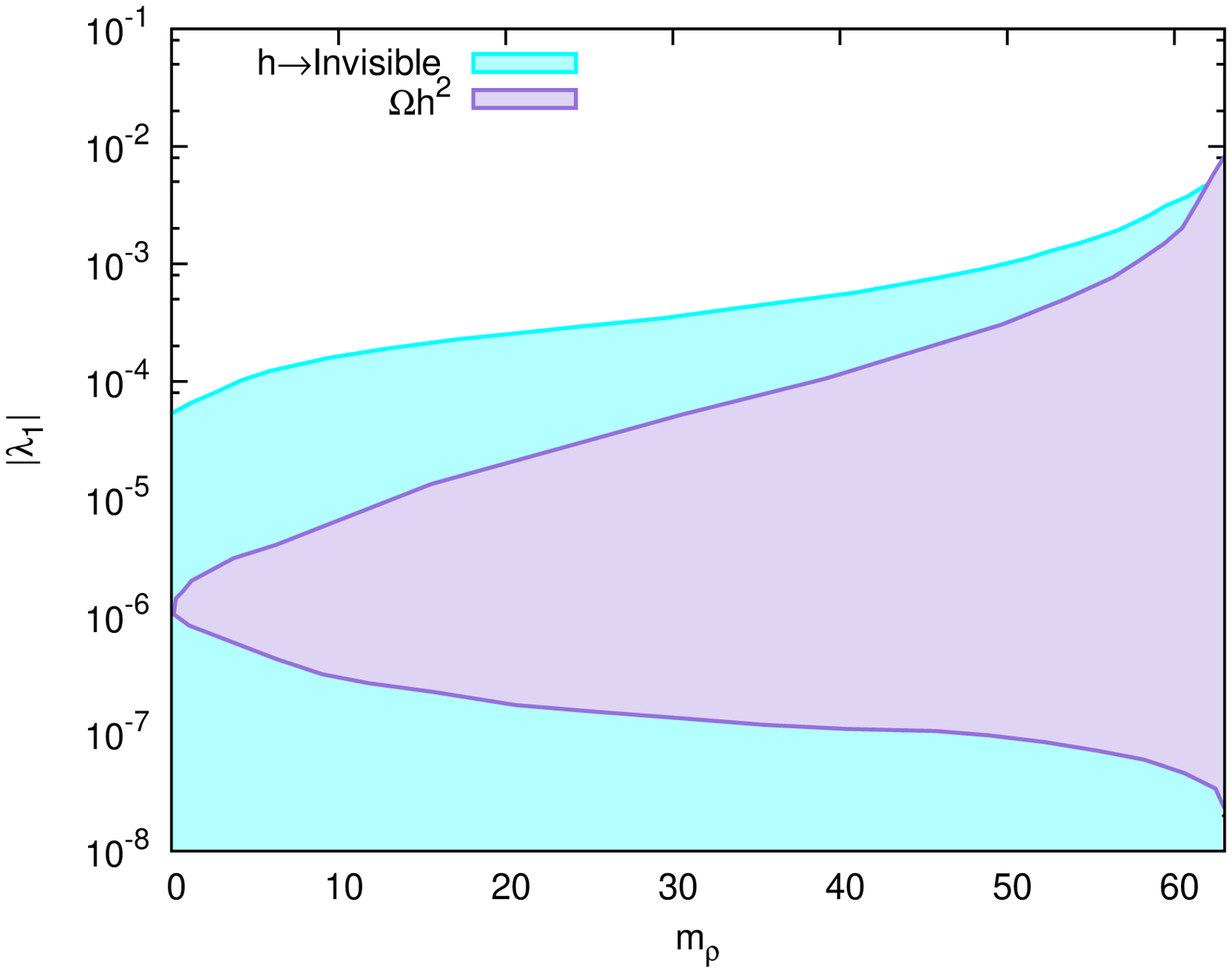,width=6.5cm}}
\centerline{\vspace{-1.5cm}\hspace{0.5cm}(a)\hspace{6cm}(b)}
\centerline{\vspace{-0.0cm}}
\end{center}
\caption{a) Shaded areas depict ranges of parameters space in mass of pseudo-scalar and  mixing coupling $|\lambda_1|$  plane which are consistent with experimental measurements of $Br(h\rightarrow \rm Invisible)$ for $m_{\rho}<62.5$  and different values of $\lambda=1, 0.1, 0.01$  b) Allowed ranges of parameters for $Br(h\rightarrow \rm Invisible)$ in comparison to regions which $\rho$  contributes more than $10\%$ to DM for $\lambda=0.01$.}\label{scaterBr}
\end{figure}

In Fig.~\ref{scaterBr}, we assume  $m_{\rho}<m_h/2$,  and illustrate regions of parameters space in $m_{\rho}$ and mixing coupling $\lambda_1$ plane which are consistent with experimental measurements $Br(h\rightarrow \rm Invisible)$. In Fig~.\ref{scaterBr}-b, we suppose $g_p=0$ and $v_{\varphi}=0$. In this case, $\rho$ play role of scalar DM. We show allowed regions in parameters space which are consistent with invisible Higgs decay and relic density of $\rho$ as DM.

\subsection{DM self-interaction}
The self-interaction of DM particles is not detectable by particle colliders or direct detection experiment and thus its coupling is not constrained by common experiments. The tightest constraint on non-gravitational interactions of DM caused from studying collision of giant clusters of 1E0657-56, so called as Bullet cluster\cite{Markevitch:2003at}.  The infalling gas of SM particles inside merging galaxies is subjected to ram pressure and lag behind the DM  \cite{Clowe:2003tk}. Using combination of weak-lensing, optical and X-ray imaging to measure this lag, has provided an upper limit on $\sigma_{DM}/m < 1.25~cm^2/g~ (68 \%~ \rm C.L)$. A recent article published in Science, studied 72 collision including minor and major mergers and claimed upper limit of $\sigma_{DM}/m < 0.47 ~cm^2/g ~(95 \% ~\rm C.L)$ \cite{Harvey:2015hha}.

Another interesting case is four elliptical galaxies in core of Abell 3827. At least one of these galaxies has a halo which is spatially offset from its stars by 1.6 Kpc \cite{Massey:2015dkw}. Noted that sole interpretation of this offset by self-interaction of DM is possible only if $\sigma_{DM}/m \sim (1.7 \pm 0.7)\times 10^{-4}~cm^2/g $. However another group \cite{Kahlhoefer:2015vua}, doubted this result because of the questionable assumptions in adoption of the model by Williams-Saha \cite{Williams:2011pm} to estimate effects of self interaction. The corrected estimation for cross-section is $\sigma_{DM}/m \sim 1.5~cm^2/g$ which slightly crosses over the previously known upper bound from Bullet cluster.\\
 So far we have presented the model of non-thermal singlet pseudo-scalar with fermionic DM and the special case of $g_p=0$ and $v_{\varphi}=0$ which is practically the singlet scalar model. In  \cite{Campbell:2015fra}, it has been shown that the model of a singlet scalar with its self-interaction within interval of $\sigma_{DM}/m \sim [1,1.5]~cm^2/g$, would only produce the sufficient relic density of DM through freeze-in mechanism. In the following, we'll study the parameter space of our model and its capacity to produce sufficiently abundant self-interacting DM. To study this bound, we have calculated, self interaction of pseudo scalar process  $\rho\rho\rightarrow \rho\rho$ (cross-section is available in Appendix) and applied DM self-interacting bound from \cite{Campbell:2015fra}.
\begin{figure}
\begin{center}
\centerline{\hspace{0cm}\epsfig{figure=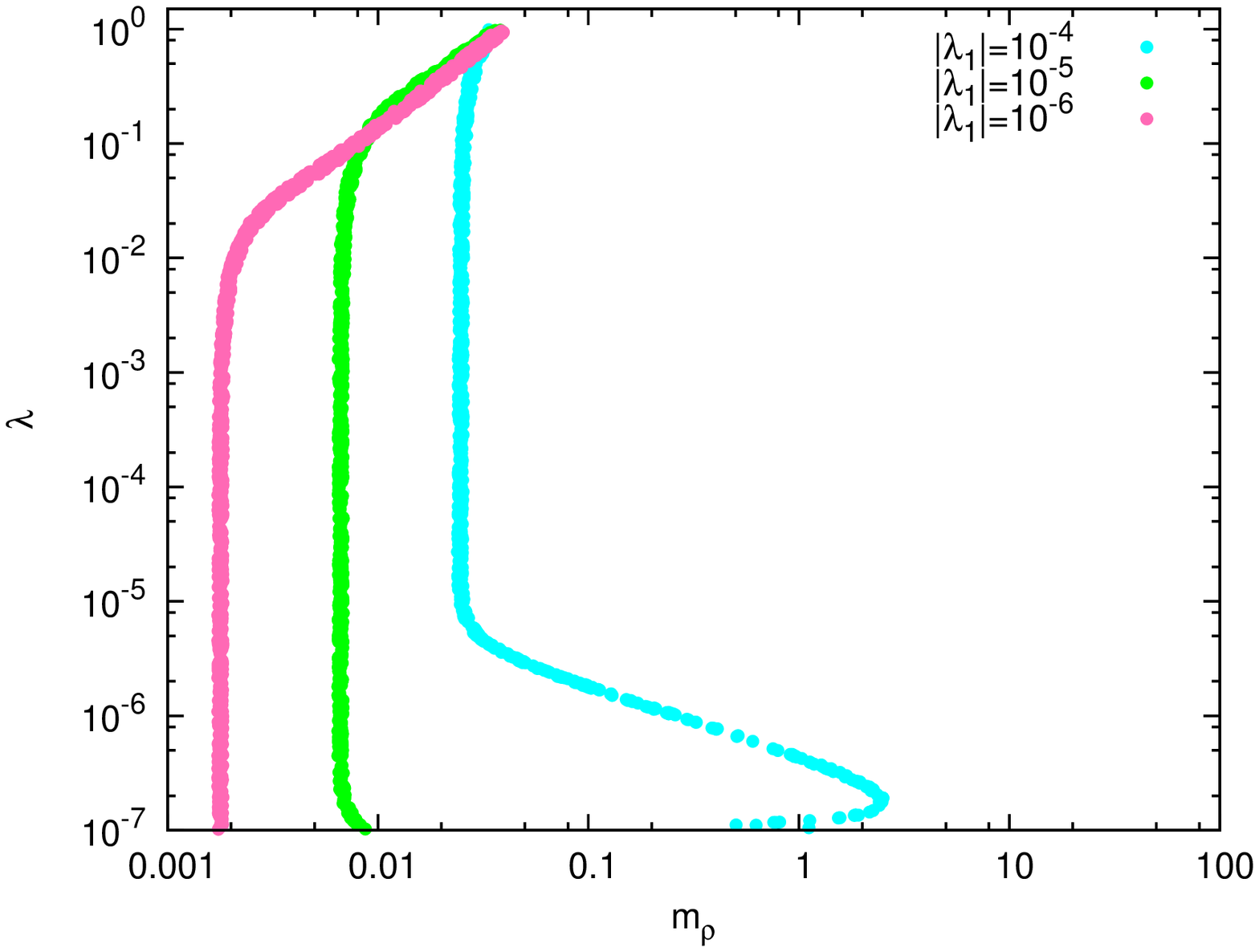,width=6.5cm}\hspace{0cm}\epsfig{figure=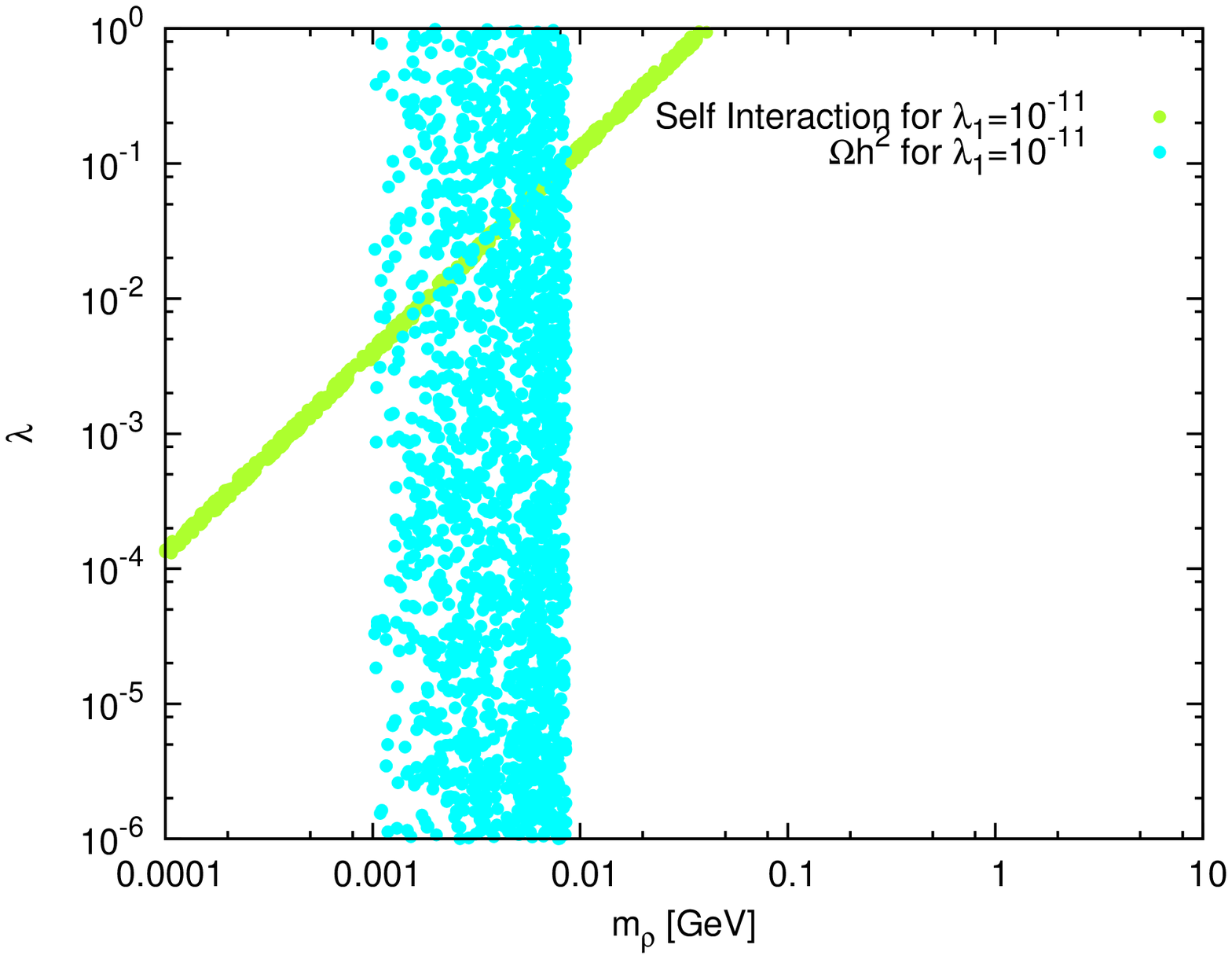,width=6.5cm}}
\centerline{\vspace{-1.5cm}\hspace{0.5cm}(a)\hspace{6cm}(b)}
\centerline{\vspace{-0.0cm}}
\end{center}
\caption{a) Colored areas depict ranges of parameters space in mass of pseudo-scalar and coupling $|\lambda|$  plane for which are consistent with  bound of self-interacting DM in the cluster Abell 3827  for different values of $\lambda_1$. b) Allowed ranges of parameters space suitable for self-interacting DM scenario overlap to regions which $\rho$  contributes more than 10 percent to DM.}\label{scaterself}
\end{figure}
In Fig.~\ref{scaterself}-a, we have shown, allowed regions in coupling $|\lambda|$  and $m_{\rho}$ plane which are consistent with  bound of self-interacting DM in the cluster Abell 3827  for different values of $\lambda_1$. As a result of this strong bounds on self-interaction, valid regions are limited to small areas.
In Fig.~\ref{scaterself}-b, we probed self-interaction and relic density constraints. As it is seen in this figure, desirable scenario for scalar DM for this case is very light DM (1-10 Mev) in which relic density is produced by non-thermal mechanism ($\lambda_1\simeq10^{-11}$). The overlapping regions in this figure express regions of parameters space which simultaneously produce significant amount of DM (greater than 10 percent of DM relic density) and justify observations of merging galaxies by self-interacting DM. Note that our results are consistent with \cite{Campbell:2015fra}.

\subsection{Indirect detection}
Recently, an excess of high energy positron has been observed by AMS-02 in milky way Galaxy \cite{AMS}. The obtained spectrum can both be explained by the annihilation of DM particles \cite{positron} or astrophysical sources \cite{pulsar}. In this section, we perform a statistical analysis of the AMS-02, positron flux, in the context of our model. We suppose  that $g_p$ and $v_{\varphi}$ are zero and consider $\rho$ as DM and study annihilation of pseudo scalar in the center of our Galaxy. We indicate how annihilation of DM particles $\rho$ will influence on estimates of $e^+$ flux and $e^+/e^-+e^+$ at the earth atmosphere measured by AMS-02.

The background flux consists of primary electrons and secondary electrons/positrons which propagate in the galaxy until they reach the earth. For this purpose, we adopted Galprop which considers different astronomical variables to calculate propagation of charged particles and cosmic rays through galaxy \cite{Indirect}. We also employed PPPC4DM package to calculate contribution of DM annihilation electron-positron flux. For every point in model parameter space, $\rho$ annihilates in various channels including $\rho\rho$ to $hh,t\bar{t},b\bar{b}, WW,ZZ$. The corresponding annihilation cross-sections are available in Appendix. The final flux consists of  background and DM flux. We will compare our predicted flux with the latest data of positron flux from AMS-02 \cite{AMS}.

\begin{figure}
\begin{center}
\centerline{\hspace{0cm}\epsfig{figure=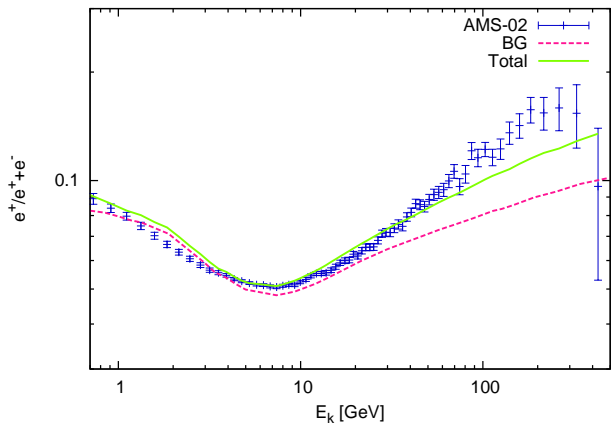,width=8.5cm}}
\centerline{\vspace{-1.3cm}}
\end{center}
\caption{ The curves show positron fraction observed by AMS-02, the background cosmic ray flux and the total $e^+$ fraction flux with DM contribution. We selected parameter values for our model with best fit of AMS-2 data.}\label{AMSfit}
\end{figure}
We perform a $\chi^2$-fit for $e^+$ fraction as a function of positron kinetic energy. This quantity has been defined as:
\begin{eqnarray}
\chi^2=\sum_{i}[\frac{(f^{th})_i-(f^{AMS})_i}{(\sigma^{AMS})_i}]^2,\label{Xai}
\end{eqnarray}
where summation is over all kinetic energy points (72 points). In Fig.~\ref{AMSfit}, we consider large values for $\lambda_1$ in which $\rho$ produces DM relic density as WIMP. Blue dots with error bars, depict positron fraction observed by AMS-02. The dashed red curve shows the background cosmic ray flux and green curve shows the total $e^+$ fraction flux with DM contribution. We selected parameter values for our model which is best fitted to AMS-2 data. As it is seen, our results indicate that addition of positron flux from $\rho$ annihilation is remarkable improvement in fitting the data. The $\chi^2$  for background is $18.72$ while  the best fit of our model obtains $\chi^2=3.3$. However it should be reminded that the calculated flux is calculated upon the assumption that the model particles produce whole the required DM in the Milky way indicated by the common density profiles. Our analysis of relic density in these regions shows they will produce DM several order of magnitude larger than the current abundance. Thus our model will not represent a reliable explanation to AMS-02 observations.
\section{Concluding Remarks}

We studied a model of non-thermal DM postulating additional fermionic field $\chi$ and the pseudo scalar $\rho$ which interact with themselves and $\rho$ mixes with SM Higgs bosons. The potential for the scalar fields of this model, has minimal terms since CP conservation preclude the scalar terms of odd power such as, $\varphi$, $\varphi^3$ and $\varphi H^2$. Motivated by the lack of signal detection in Xenon100 and LUX experiments, we considered regimes of very small coupling between fermionic DM and pseudo scalar fields which produce DM via non-thermal freeze-in mechanism.  However pseudo scalar $\rho$ can reach equilibrium when there is a large enough coupling with Higgs field. In section.$~3.1$, we studied DM production in which two newly added fields $\chi$ and $\rho$ are FIMP. In this case, we have shown that for very small coupling $g_p$ and $\lambda_1$ Boltzmann equation  for fermionic DM is independent from pseudo scalar abundance.  We have discussed the allowed regions in parameters space of our model in consistency with relic density measurement.
We have shown for the case that $g_p=0$ and $v_{\varphi}=0$, there is no mixing between Higgs field and pseudo scalar field and as a result, $\rho$ abundance remains constant and the stable  $\rho$ plays role of DM.

We also studied DM production in the case of  fermionic DM as FIMP and pseudo scalar particle $\rho$ enters thermal equilibrium so act as a WIMP. Yield quantity and abundance of new fields as a function of temperature have been studied in secrion.~$3.2$. In this case, different processes may contribute to the DM production. It was shown that the main contribution to DM abundance arises from annihilation of $\rho\rho$ and other processes such as annihilation of $WW$, $ZZ$, $hh$ or $bb$ are subdominant. Also it is shown for the case in which $\rho$ is produced thermally and $g_p=0$ and $v_{\varphi}=0$, pseudo can play role of DM and act as a WIPM.

 We have also found viable regions in parameters space in agreement with new upper limit on invisible Higgs decay branching ratio. In section.~$4$, we compared consistent region in parameters space for invisible Higgs decay with relic density of pseudo scalar DM. 

Moreover, in regimes in which $g_p$ and $v_{\varphi}$ are zero $\chi$ becomes irrelevant and $\rho$ will claim the role for DM.
 For this case, we mentioned recent observations of merging galaxies and illustrated regions of parameters space in which self-interaction of DM particles are strong enough to explain these observations while the DM is abundant. We also performed statistical analysis of AMS-02 positron fluxes in the context of our model. We showed that adding pseudo scalar $\rho$ DM contribution improves the fit to the AMS-02 data.

\section{Acknowledgement}
We would like to thank A. Merle, M. Torabian and M. Tavakoli for the useful discussions and comments.
\section{Appendix}
In this appendix, we summarize the formulae of production cross sections which contribute to relic density abundance of fermionic DM in our model.
The production cross section of DM pair from SM fermions is given by:
\label{sec:two}
\unboldmath
\begin{eqnarray}
\sigma_{pro}v_{rel}(\overline{f}f\rightarrow\overline{\chi}\chi)&=&\frac{g_p^2\sin^2{2}\theta}{32\pi}[N_cs(\frac{m_f}{v_0})^2(1-\frac{4m_f^2}{s})(1-\frac{4m_\chi^2}{s})^{1/2}]
\nonumber\\ &\times&[\frac{1}{(s-m_h^2)^2+m_h^2\Gamma_h^2}+\frac{1}{(s-m_\rho^2)^2+m_\rho^2\Gamma_\rho^2}\nonumber\\ &-&\frac{2(s-m_h^2)(s-m_\rho^2)+2m_hm_\rho\Gamma_h\Gamma_\rho}{((s-m_h^2)^2+m_h^2\Gamma_h^2)((s-m_\rho^2)^2+m_\rho^2\Gamma_\rho^2)}],
\end{eqnarray}
which $N_c$ is the number of color charge. The total cross section of DM from SM gauge bosons ($Z$ and $W^{\pm}$) are given by:
\begin{eqnarray}
\sigma_{pro}v_{rel}(VV\rightarrow\overline{\chi}\chi)&=&\frac{g_p^2\sin^2{2}\theta}{72\pi}\times[\frac{m_V^4}{v_0^2}(2+\frac{(s-2m_V^2)^2}{4m_V^4})(1-\frac{4m_\chi^2}{s})^{1/2}]
\nonumber\\&\times&[\frac{1}{(s-m_h^2)^2+m_h^2\Gamma_h^2}
+\frac{1}{(s-m_\rho^2)^2+m_\rho^2\Gamma_\rho^2}\nonumber\\
&-&\frac{2(s-m_h^2)(s-m_\rho^2)+2m_hm_\rho\Gamma_h\Gamma_\rho}{((s-m_h^2)^2+m_h^2\Gamma_h^2)((s-m_\rho^2)^2+m_\rho^2\Gamma_\rho^2)}],
\end{eqnarray}
and we also calculate the following formula for production of DM from two neutral Higgs-like  scalar $h$ and $\rho$:
\begin{eqnarray}
&&\sigma_{pro}v_{rel}(hh\rightarrow\overline{\chi}\chi)=\frac{g_p^2}{4\pi}(1-\frac{4m_\chi^2}{s})^{1/2}[\frac{a^2\sin^2\theta}{(s-m_h^2)^2+m_h^2\Gamma_h^2}\nonumber\\&+&
\frac{b^2\cos^2\theta}{(s-m_{\rho}^2)^2+m_{\rho}^2\Gamma_{\rho}^2}+\frac{ab\sin2\theta((s-m_h^2)(s-m_\rho^2)+m_h m_\rho\Gamma_h\Gamma_\rho)}{((s-m_h^2)^2+m_h^2\Gamma_h^2)((s-m_{\rho}^2)^2+m_{\rho}^2\Gamma_{\rho}^2)}]\nonumber\\
&+&\frac{g_p^4\sin^4\theta}{\pi s}(1-\frac{4m_\chi^2}{s})^{1/2} \frac{1}{(s-2m_h^2)^2 x_1(x_1^2-1)}
\times[x_1(6m_h^4-4m_h^2 s+s^2\nonumber\\&-&(s-2m_h^2)^2 x_1^2)+(6m_h^4-4m_h^2 s+s^2)(x_1^2-1)arctanh(x_1)],
\end{eqnarray}
where $x_1=\frac{(s-4m_\chi^2)^\frac{1}{2} (s-4m_h^2)^\frac{1}{2}}{(s-2m_h^2)}$,
\begin{eqnarray}
a&=&-\lambda v_\varphi\sin^3\theta +6\lambda_H v_H\cos^3\theta -6\lambda_1 v_H\sin^2\theta\cos\theta -6\lambda_1 v_\varphi\cos^2\theta\sin\theta\,  ,\nonumber\\
b&=&-\lambda v_\varphi\cos\theta\sin^2\theta -6\lambda_H v_H\cos^2\theta\sin\theta+2\lambda_1 v_H\sin\theta+6\lambda_1 v_\varphi\cos\theta\sin^2\theta\nonumber\\
&-& 2\lambda_1 v_\varphi\cos\theta-6\lambda_1 v_H\cos^2\theta\sin\theta \, ,\nonumber
\end{eqnarray}
 and production cross section from two $\rho$ is:
\begin{eqnarray}
&&\sigma_{pro}v_{rel}(\rho\rho\rightarrow\overline{\chi}\chi)=\frac{g_p^2}{4\pi}(1-\frac{4m_\chi^2}{s})^{1/2}[\frac{c^2\sin^2\theta}{(s-m_h^2)^2+m_h^2\Gamma_h^2}\nonumber\\
&+&\frac{d^2\cos^2\theta}{(s-m_\rho^2)^2+m_\rho^2\Gamma_\rho^2}+\frac{cd\sin2\theta((s-m_h^2)(s-m_\rho^2)+m_hm_\rho\Gamma_h\Gamma_\rho)}{((s-m_h^2)^2+m_h^2\Gamma_h^2)((s-m_\rho^2)^2+m_\rho^2\Gamma_\rho^2)}]\nonumber\\
&+&\frac{g_p^4\cos^4\theta}{\pi s}(1-\frac{4m_\chi^2}{s})^{1/2}
\frac{1}{(s-2m_\rho^2)^2 x_2(x_2^2-1)}\times
[x_2(6m_\rho^4-4m_\rho^2 s+s^2\nonumber\\&-&(s-2m_\rho^2)^2 x_2^2)+(6m_\rho^4-4m_\rho^2 s+s^2)(x_2^2-1)arctanh(x_2)],
\end{eqnarray}
with
\begin{eqnarray}
c&=&-\lambda v_\varphi\cos^2\theta\sin\theta+6\lambda_H v_H\sin^2\theta\cos\theta+6\lambda_1 v_H\sin^2\theta\cos\theta-2\lambda_1 v_H\cos\theta\nonumber\\
&-& 2\lambda_1 v_\varphi\sin\theta+6\lambda_1 v_\varphi\cos^2\theta\sin\theta  ,\nonumber\\
d&=&-\lambda v_\varphi\cos^3\theta-6\lambda_H v_H\sin^3\theta +6\lambda_1 v_H\cos^2\theta\sin\theta -6\lambda_1 v_\varphi\sin^2\theta\cos\theta   ,\nonumber
\end{eqnarray}
where $x_2=\frac{(s-4m_\chi^2)^\frac{1}{2} (s-4m_\rho^2)^\frac{1}{2}}{(s-2m_\rho^2)}$.
In our model, there is another contributions for DM density from annihilation of $h$ and $\rho$ into DM pair. The cross section for this process is expressed by:
\begin{eqnarray}
&&\sigma_{pro}v_{rel}(h\rho\rightarrow\overline{\chi}\chi)=\frac{g_p^2}{4\pi}(1-\frac{4m_\chi^2}{s})^{1/2}[\frac{b^2\sin^2\theta}{(s-m_h^2)^2+m_h^2\Gamma_h^2}\nonumber\\
&+&\frac{c^2\cos^2\theta}{(s-m_\rho^2)^2+m_\rho^2\Gamma_\rho^2}+\frac{bc\sin2\theta((s-m_h^2)(s-m_\rho^2)+m_hm_\rho\Gamma_h\Gamma_\rho)}{((s-m_h^2)^2+m_h^2\Gamma_h^2)((s-m_\rho^2)^2+m_\rho^2\Gamma_\rho^2)}]\nonumber\\
&+&\frac{g_p^4\sin^22\theta}{2\pi s}(1-\frac{4m_\chi^2}{s})^{1/2}\frac{1}{(s-2m_\rho^2)^2 x_3(x_3^2-1)}\times[x_3(2m_h^2m_\rho^2\nonumber\\&-&(s-2m_\rho^2)^2(x_3^2-1))+(2m_h^2m_\rho^2+(s-2m_\rho^2)^2)(x_3^2-1)arctanh(x_3)]     ,\nonumber\\
\end{eqnarray}
where $x_3=\frac{(s-4m_\chi^2)^\frac{1}{2} (-2m_h^2-2m_\rho^2+s)^\frac{1}{2}}{(s-2m_\rho^2)}$.
We also obtain annihilation cross sections of pseudo scalar pair into a pair of SM model particles $h, W, Z$ and $f$  as:
\begin{eqnarray}
&&\sigma_{ann}v_{rel}(\rho\rho\rightarrow hh)=\frac{1}{8\pi s}(1-\frac{4m_h^2}{s})^{1/2}[\frac{a^2c^2}{(s-m_h^2)^2+m_h^2\Gamma_h^2}\nonumber\\&+&\frac{b^2d^2}{(s-m_\rho^2)^2+m_\rho^2\Gamma_\rho^2}
+\frac{2abcd((s-m_h^2)(s-m_\rho^2)+m_hm_\rho\Gamma_h\Gamma_\rho)}{((s-m_h^2)^2+m_h^2\Gamma_h^2)((s-m_\rho^2)^2+m_\rho^2\Gamma_\rho^2)}]\nonumber,\\
\end{eqnarray}
\begin{eqnarray}
&&\sigma_{ann}v_{rel}(\rho\rho\rightarrow W^+W^-,ZZ)=\frac{1}{8\pi s}[\frac{c^2\cos^2\theta}{(s-m_h^2)^2+m_h^2\Gamma_h^2}+\frac{d^2\sin^2\theta}{(s-m_\rho^2)^2+m_\rho^2\Gamma_\rho^2}\nonumber\\
&-&cd\sin2\theta\frac{(s-m_h^2)(s-m_\rho^2)+m_hm_\rho\Gamma_h\Gamma_\rho}{((s-m_h^2)^2+m_h^2\Gamma_h^2)((s-m_\rho^2)^2+m_\rho^2\Gamma_\rho^2)}]\times\nonumber\\
&[&4(\frac{m_W^2}{v_H})^2(2+\frac{(s-2m_W^2)^2}{4m_W^4})(1-\frac{4m_W^2}{s})^{1/2} +2(\frac{m_Z^2}{v_H})^2(2+\frac{(s-2m_Z^2)^2}{4m_Z^4})(1-\frac{4m_Z^2}{s})^{1/2}]\nonumber,\\
\end{eqnarray}
\begin{eqnarray}
&&\sigma_{ann}v_{rel}(\rho\rho\rightarrow\overline{f}f)=\frac{1}{32\pi s}(\frac{m_f}{v_H})^2(1-\frac{4m_f^2}{s})^{1/2}[2(s-4m_f^2)\nonumber\\&\times&(\frac{c^2\cos^2\theta}{(s-m_h^2)^2+m_h^2\Gamma_h^2}+\frac{d^2\sin^2\theta}{(s-m_\rho^2)^2+m_\rho^2\Gamma_\rho^2}
\nonumber\\&-&\frac{cd\sin2\theta((s-m_h^2)(s-m_\rho^2)+m_hm_\rho\Gamma_h\Gamma_\rho)}{((s-m_h^2)^2+m_h^2\Gamma_h^2)((s-m_\rho^2)^2+m_\rho^2\Gamma_\rho^2)})\nonumber\\
&+&\frac{m_f}{v_H}\sin^2\theta(\frac{c\cos\theta (s-m_h^2)}{(s-m_h^2)^2+m_h^2\Gamma_h^2}-\frac{d\sin\theta (s-m_\rho^2)}{(s-m_\rho^2)^2+m_\rho^2\Gamma_\rho^2})\nonumber\\
&\times& 32m_f(1+\frac{(s-8m_f^2+2m_\rho^2)ArcTanh(x_4)}{(s-2m_\rho^2)x_4})\nonumber\\
&+&(\frac{m_f}{v_H})^2\sin^4\theta\frac{16}{(s-2m_\rho^2)^2x_4(x_4^2-1)}(2x_4(m_\rho^2-4m_f^2)^2\nonumber\\
&-&(s-2m_\rho^2)^2x_4(x_4^2-1)+((s-2m_\rho^2)^2\nonumber\\
&+&16m_f^2(s-m_\rho^2-2m_f^2)+2m_\rho^4)(x_4^2-1)arctanh(x_4))]\nonumber.\\
\end{eqnarray}
where $x_4=\frac{(s-4m_f^2)^\frac{1}{2} (s-4m_\rho^2)^\frac{1}{2}}{(s-2m_\rho^2)}$.
Self-interacting cross section for process $\rho\rho\rightarrow \rho\rho$ is given by ($\theta=0$):
\begin{eqnarray}
&&\sigma(\rho\rho\rightarrow \rho\rho)=\frac{1}{16\pi s}(1-\frac{4m_\rho^2}{s})^{1/2}[\frac{c^4}{(s-m_h^2)^2+m_h^2\Gamma_h^2}\nonumber\\
&+&\frac{d^4}{(s-m_\rho^2)^2+m_\rho^2\Gamma_\rho^2}+\frac{2c^2d^2((s-m_h^2)(s-m_\rho^2)+m_hm_\rho\Gamma_h\Gamma_\rho)}{((s-m_h^2)^2+m_h^2\Gamma_h^2)((s-m_\rho^2)^2+m_\rho^2\Gamma_\rho^2)}\nonumber\\
&+&\frac{\lambda}{2}m_h\Gamma_h \frac{c^2\cos^4\theta }{(s-m_h^2)^2+m_h^2\Gamma_h^2}+\frac{\lambda}{2}m_\rho\Gamma_\rho\frac{d^2\cos^4\theta }{(s-m_\rho^2)^2+m_\rho^2\Gamma_\rho^2}\nonumber\\
&+&\frac{\lambda^2}{16}\cos^8\theta].
\nonumber\\
\end{eqnarray}

\end{document}